\RequirePackage{rotating}
\documentclass[acmsmall]{acmart}

\def\BibTeX{{\rm B\kern-.05em{\sc i\kern-.025em b}\kern-.08emT\kern-.1667em\lower.7ex\hbox{E}\kern-.125emX}}

\usepackage[ruled]{algorithm2e} % For algorithms

\usepackage[symbol]{footmisc}

\SetAlFnt{\small}
\SetAlCapFnt{\small}
\SetAlCapNameFnt{\small}
\SetAlCapHSkip{0pt}
\IncMargin{-\parindent}

\usepackage{multirow}
\usepackage{tabu}
\usepackage{multicol}
\usepackage{tabularx}
\usepackage{color, colortbl}
\definecolor{Gray}{gray}{0.9}
\usepackage{hhline}
\usepackage{rotating}
\usepackage{lscape}
\usepackage{graphicx}
\usepackage{subfig}
\acmJournal{CSUR}
\acmVolume{53}
\acmNumber{3}
\acmArticle{1}
\acmYear{2020}
\acmMonth{4}
\copyrightyear{2020}
\begin{document}

	% The "title" command has an optional parameter, allowing the author to define a "short title" to be used in page headers.
	\title{Security {and} Privacy in IoT Using Machine Learning {and} Blockchain: Threats {and} Countermeasures}

	% The "author" command and its associated commands are used to define the authors and their affiliations.
	% Of note is the shared affiliation of the first two authors, and the "authornote" and "authornotemark" commands used to denote shared contribution to the research.
	
	\thanks{This research is supported by the Australian Government Research Training Program Scholarship.}
	\author{Nazar Waheed}
	
	\affiliation{%
		\streetaddress{School of Electrical and Data Engineering}
		\institution{University of Technology Sydney}
		\postcode{2007}
		\city{Sydney}
		\state{New South Wales}
		\country{Australia}}
	\email{nazar.waheed@student.uts.edu.au}
	
	\author{Xiangjian He*}
	\thanks{*Corresponding author.}
	\affiliation{%
		\streetaddress{School of Electrical and Data Engineering}
		\institution{University of Technology Sydney}
		\postcode{2007}
		\city{Sydney}
		\state{New South Wales}
		\country{Australia}}
	\email{xiangjian.he@uts.edu.au}
	
	\author{Muhammad Ikram}
	\affiliation{%
		\streetaddress{Department of Computing}
		\institution{Macquarie University}
		\postcode{2109}
		\city{North Ryde}
		\state{New South Wales}
		\country{Australia}}
	\email{muhammad.ikram@mq.edu.au}
	
	\author{Muhammad Usman}
	\affiliation{%
		\streetaddress{School of Computing and Mathematics}
		\institution{University of South Wales}
		\postcode{CF37 1DL}
		\city{Pontypridd}
		\state{Rhondda Cynon Taff}
		\country{United Kingdom}}
	\email{muhammad.usman@southwales.ac.uk}
	
	\author{Saad Sajid Hashmi}
	\affiliation{%
		\streetaddress{Department of Computing}
		\institution{Macquarie University}
		\postcode{2109}
		\city{North Ryde}
		\state{New South Wales}
		\country{Australia}}
	\email{saad.hashmi@hdr.mq.edu.au}
	
	\author{Muhammad Usman}
	\affiliation{
		\streetaddress{School of Science, Engineering and IT}
		\institution{Federation University}
		\postcode{3350}
		\city{Mt Hellen}
		\state{Victoria}
		\country{Australia}}
	\email{muhammad.usmanskk@gmail.com}

	\renewcommand{\shortauthors}{Nazar et al.}

	% The abstract is a short summary of the work to be presented in the article.
	\begin{abstract}
		Security and privacy {of the users} have become significant concerns due to the involvement of the Internet of Things (IoT) devices in {numerous} 
		applications. Cyber threats are growing at an explosive pace making the existing security and privacy measures inadequate. Hence, everyone on the Internet is a product for hackers. {Consequently, Machine Learning (ML) algorithms are used to produce accurate outputs from large complex databases, where the generated outputs can be used to predict and detect vulnerabilities in IoT-based systems. Furthermore, Blockchain (BC) techniques are becoming popular in modern IoT applications to solve security and privacy issues.} Several studies have been conducted on either ML algorithms or BC techniques. However, these studies target either security or privacy issues using ML algorithms or BC techniques, thus posing a need for a combined survey on efforts made in recent years addressing both security and privacy issues using ML algorithms and BC techniques. {In this paper, we provide a summary of research efforts made in the past few years, starting from 2008 to 2019, addressing security and privacy issues using ML algorithms and BC techniques in the IoT domain.} {First, we discuss and categorize various security and privacy threats reported in the past twelve years in the IoT domain. We then classify the literature on security and privacy efforts based on ML algorithms and BC techniques in the IoT domain. Finally, we identify and illuminate several challenges and future research directions using ML algorithms and BC techniques to address security and privacy issues in the IoT domain.}
		
	\end{abstract}
	
	
	
	\terms{Attack, Vulnerability, Algorithm}
	
	
	\maketitle
	
	%%%%% Introduction %%%%%
	\vspace{-0.5cm}
	\section{Introduction}
	
	We have seen the industries to evolve from manufacturing just the \textit{products} to building the \textit{network of products} known as the \textit{Internet of Things} (IoT), and eventually creating an intelligent \textit{network of products} {providing various, invaluable online service \cite{Dalgleish2007, rawat2017industrial}}. As per Aksu et al. \cite{Aksu2018}, two devices are connected to the Internet every three minutes. This connectivity and the exponential growth of IoT devices have resulted in an increased amount of network traffic. Due to this connectivity, challenges like security and privacy of user data and verification and authentication of devices, have arisen~\cite{song2017security}. For example, hackers compromised one billion yahoo accounts in 2013 \cite{Goel2016}. In 2014, one hundred and forty-five million eBay users were under attack \cite{Peterson2014}. Following the increasing trend of attacks, in 2017, one hundred and forty-three million customers from Equifax had their personal information stolen \cite{Zhou2018}.
	Similarly, as reported in \cite{Kshetri2017}, a five billion dollar toy industry in 2017 had their eight hundred and twenty thousand client accounts compromised. It also included over two million voice recordings, out of which a few were held for ransom. The recent cyber history is full of cybersecurity disasters, from massive data breaches to security flaws in billions of microchips and computer system lockdowns until a payment was made \cite{Giles2019}. There are a plethora of security and privacy challenges for IoT devices, which are increasing every day. Hence, security and privacy in complex and resource-constrained IoT environments are big challenges and need to be tackled effectively.
	
	The security challenges in IoT are increasing as the attacks are getting sophisticated day by day. Milosevic et al. \cite{Milosevic2016} highlighted that powerful computing devices, e.g., desktop computers, might be able to detect malware using sophisticated resources. However, IoT devices have limited resources. Similarly, traditional cybersecurity systems and software are not efficient enough in detecting small attack variations or zero-day attacks \cite{Chaabouni2019}, since both need to be updated regularly. Moreover, the updates are not available by the vendor in real-time, making the network vulnerable. {Machine Learning (ML) algorithms can be employed to improve IoT infrastructure (such as smart sensors and IoT gateways)~\cite{dartmann2019big},  and also to improve the performance of cybersecurity systems~\cite{Xiao2018}}. Based on the existing knowledge of cyber-threats, these algorithms can analyze network traffic, update threat knowledge databases, and keep the underlying systems protected from new attacks \cite{Aksu2018,Xiao2016,Xiao2018}. Alongside using ML algorithms, the researchers have also started using revolutionary Blockchain (BC) technique to protect the underlying systems \cite{Zhou2018BC,Rahulamathavan2017,Wang2018,Li2018a,Lu2018,Lee2017_Priv,Lee2017_Sec,Fan2018_Security,Fan2018_Privacy}. Although 
	ML algorithms and BC techniques have been developed to deal with cyber threats in the IoT domain; combining these two is something new that needs to be explored.
	
	Privacy goes hand-in-hand with security. Price et al. defined privacy as an application-dependent set of rules \cite{Price2019}. The authors elaborate that the rules on how the information can flow depend on the involved entities, processes, frequency, and motives to access data. There are many applications, such as wearable devices \cite{Aksu2018}, Vehicular Area NETwork (VANET) \cite{Zhang2018}, health-care \cite{Zhu2017}, and smart-home \cite{Song_SmartHomes2018, Dorri2016, Dorri2017}, that require providing security and protecting the privacy of personal information. For example, in a crowdsensing application like VANET, the network is dependent on the data collected from devices to make intelligent decisions on the latest traffic conditions. However, the users of devices might be hesitant to participate due to inadequate privacy-preserving mechanisms and related threats. Extensive research works based on ML algorithms and BC techniques \cite{Aksu2018, Xiao2016, Xiao2018, Zhou2018BC, Rahulamathavan2017, Wang2018, Li2018a, Lu2018, Lee2017_Sec, Lee2017_Priv, Fan2018_Security, Fan2018_Privacy} have been conducted in the past few years to protect data on devices and preserve user's privacy.
	
	\begin{figure}[!ht]
		\centering
		\begin{center}
			\includegraphics[width=5.25in,height=3.75in]{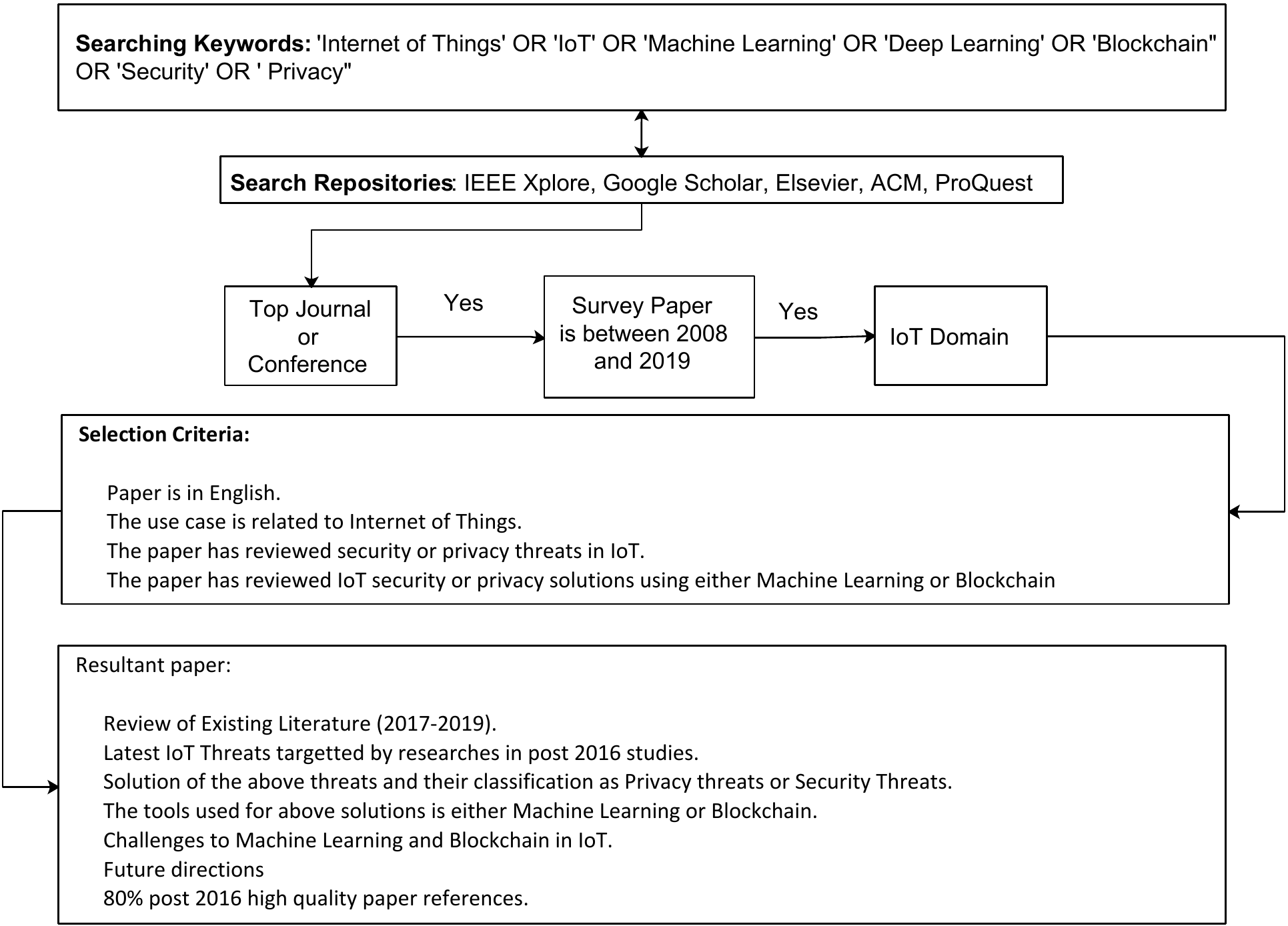}
			\vspace{-0.2cm}
			\caption{Paper collection criteria.}
			\label{fig:SelectionCriteria}
		\end{center}
		\vspace{-0.4cm}
	\end{figure}

	%Gaps in Existing Surveys
	\begin{table}[h]
		\centering
		\caption{{Contributions and gaps of all published survey papers from 2017 to 2019.}}
		\vspace{-0.2cm}
		\begin{tabular}{|l|c|c|c|c|c|}
			\hline
			\textbf{Authors [Ref.]} & \textbf{Year} & \textbf{IoT Security} & \textbf{IoT Privacy} & \textbf{Machine Learning} & \textbf{Blockchain} \\ \hline
			
			{Kshetri et al.} \cite{Kshetri2017}  & {2017}     & X   & \checkmark & X & \checkmark  \\ \hline
			
			{Banerjee et al.} \cite{Banerjee2018}    & {2018}   & \checkmark   & X& X & \checkmark \\ \hline	
			
			{Restuccia et al.} \cite{Restuccia2018} & 2018 & \checkmark   & X & \checkmark & X \\ \hline
			
			{Sharmeen et al.} \cite{Sharmeen2018}  & 2018    & \checkmark   & X & \checkmark & X \\ \hline
			
			{Xiao et al.} \cite{Xiao2018c}   & 2018    & X   & \checkmark & \checkmark & X \\ \hline

			{Khan et al.} \cite{Khan2018}   & 2018    & \checkmark   & X & X  & \checkmark \\ \hline
			
			{Reyna et al.} \cite{Reyna2018}   & 2018    & \checkmark   & X & X  & \checkmark \\ \hline
			
			{Panarello et al.} \cite{Panarello2018}  & 2018   & \checkmark   & X & X & \checkmark \\ \hline
			
			{Kumar et al.} \cite{Kumar2018}     & 2018  & \checkmark  & \checkmark & X & \checkmark   \\ \hline
			
			{Kouicem et al.} \cite{Kouicem2018}     & 2018  & \checkmark   & \checkmark & X & \checkmark   \\ \hline

			{Zhu et al. \cite{Zhu2018}}     & 2018  & X   & \checkmark & X & \checkmark   \\ \hline
			
			{Chaabouni et al.} {\cite{Chaabouni2019}} & 2019 & \checkmark & X & \checkmark & X \\ \hline
			
			{Hassija et al.} {\cite{Hassija2019}} & 2019 & \checkmark & X & \checkmark & \checkmark \\ \hline
			
			{Costa et al.} {\cite{Costa2019}} & 2019 & \checkmark & X & \checkmark & X \\ \hline
			
			{Wang et al.} {\cite{Wang2019}} & 2019 & \checkmark & X & X & \checkmark \\ \hline
			
			{Ali et al.} {\cite{Ali2019}} & 2019 & \checkmark & \checkmark & X & \checkmark \\ \hline
			
			This Survey & {2020} & \checkmark & \checkmark & \checkmark & \checkmark\\ \hline
			
		\end{tabular}
		\vspace{-0.2cm}
		\label{table:Gaps} 
	\end{table}
	
\begin{figure}
		\centering
		\subfloat[Year-wise distribution]{{\includegraphics[height=4.5cm,width=5cm]{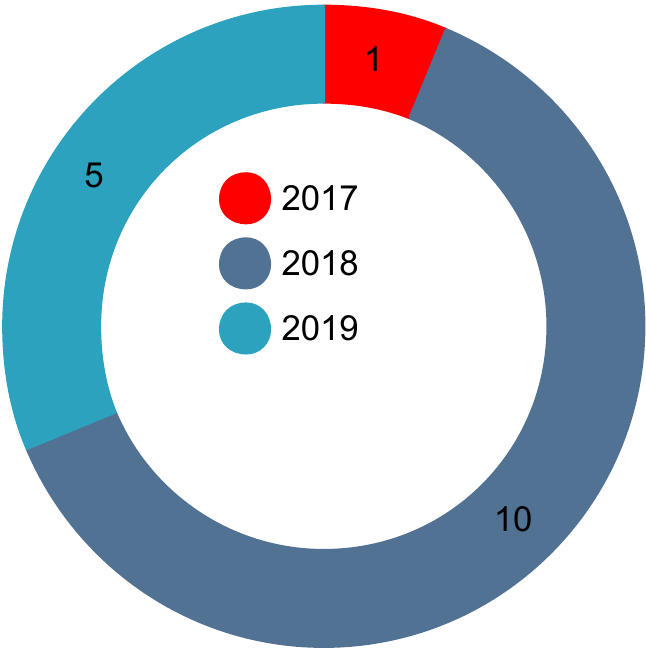} }\label{subfig2a}}
		\qquad
		\subfloat[Survey Scope]{{\includegraphics[height=4.5cm,width=5cm]{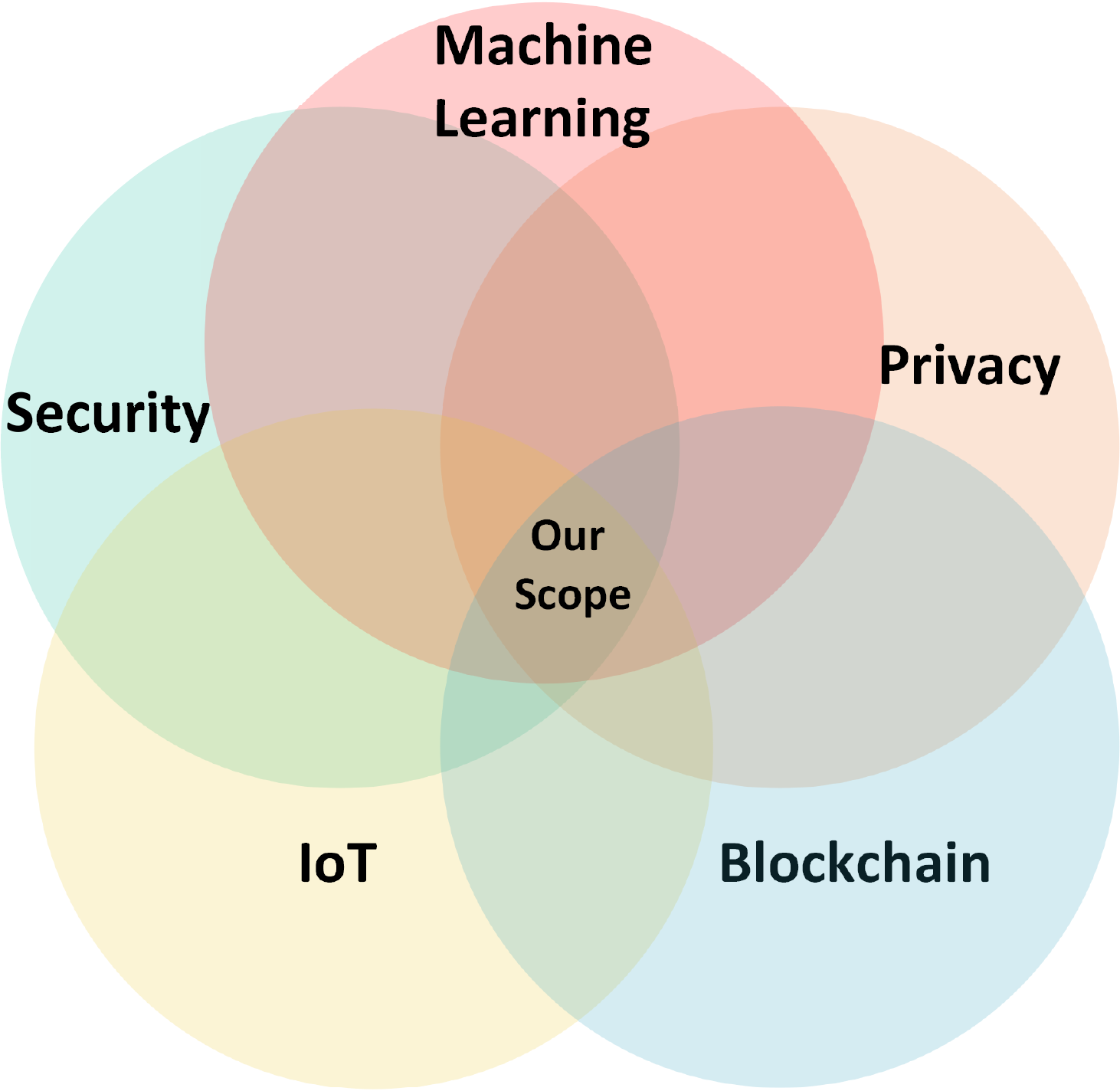} }\label{subfig2b}}

		\vspace{-0.2cm}
		\caption{{(a) Year-wise statistics of the selected survey papers between 2008 and 2019 inclusive. It shows that most of the work only started recently. (b) Scope of our study highlighting the use of ML and BC techniques to address security and privacy issues in IoT domains.}} 
		\label{fig:2in1}
		\vspace{-0.4cm}
	\end{figure}

	\textbf{Paper collection:} %The strategy of selecting articles for this study is depicted in 
	Figure \ref{fig:SelectionCriteria} depicts the strategy of selecting articles for this study. Initially, using the keywords and mentioned databases, the search was performed. The keywords such as IoT, Internet of Things, privacy, security, machine learning, and blockchain were utilized to download the latest articles from the top journals and conferences. {In order to qualify for selection, a paper must satisfy \textit{all} of the following conditions: (i) published between 2008-2019 (inclusive); (ii) be a generic (not application specific) IoT survey paper; (iii) discussed security or privacy threats related to IoT and (iv) covered ML and/or BC as a computing paradigm.} The year-wise articles selection statistics are depicted in Figure \ref{subfig2a}.
	
\begin{figure}[]
		\centering
		\begin{center}
			\vspace{-0.2cm}
			\includegraphics[width=\textwidth,keepaspectratio]{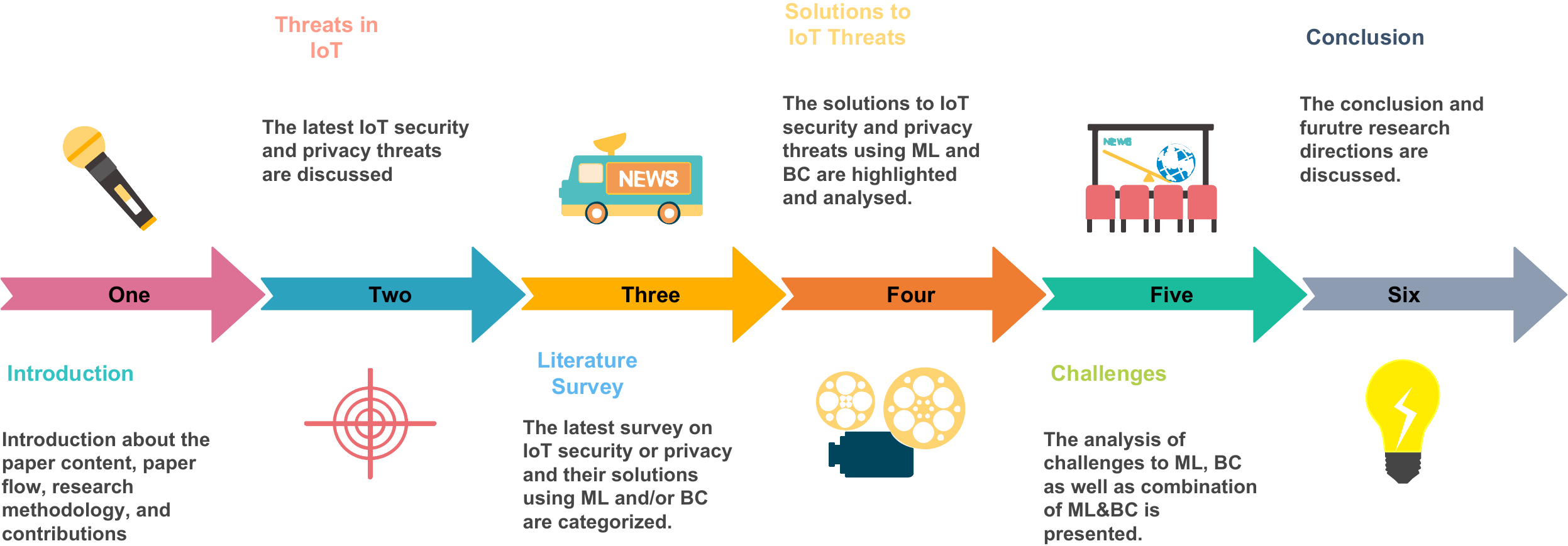}
			\vspace{-0.2cm}
			\caption{{Roadmap of our study.}}
			\label{fig:PaperFlow}
		\end{center}
		\vspace{-0.4cm}
	\end{figure}

	\textbf{Contributions of the paper:} This paper provides a detailed review of ML algorithms and BC techniques employed to protect IoT applications from security and privacy attacks. Based on the review, we highlight that a combination of ML algorithms and BC techniques can offer more effective solutions to security and privacy challenges in the IoT environment. To the best of our knowledge, this is the first paper that presents a review of security and privacy vulnerabilities in the IoT environment and their countermeasures based on ML algorithms and BC techniques. A road map of our paper is depicted in Figure \ref{fig:PaperFlow}, while Figure~\ref{subfig2b} illustrates the scope of this survey paper.

	To cover the gaps in current literature (as summarized in Table \ref{table:Gaps}), the major contributions of this paper can be summarized as follows.

	\begin{itemize}
		
		\item We provide a generic classification of IoT threats reported in recent literature based upon security and privacy threats. 
		
		\item We classify literature reviews on ML algorithms and BC techniques for IoT security and privacy, and highlight the research gaps in the existing literature reviews as in Tables \ref{table:ExistingSurveysMLSecurity}, 
		\ref{table:ExistingSurveysBCSecurity} and \ref{table:ExistingSurveysBCPrivacy}.
		
		\item We provide a taxonomy of the latest security and privacy solutions in IoT using ML algorithms and BC techniques. 
		
		\item We also identify and analyze the integration of ML algorithms with BC techniques to strengthen security and privacy in IoT.
		
		\item Finally, we highlight and discuss existing challenges to ML algorithms and BC techniques in IoT security and privacy with an attempt to suggest some future directions.
	\end{itemize} 
	
	%%%% Figure PaperFlow %%%%%
	
	The rest of this paper is organized as follows. In Section~\ref{sec:section2:threats-in-iot}, the classification of well-known IoT threats is presented. In Section~\ref{sec:section3:lit-survey}, we categorize literature reviews on IoT security and privacy using ML algorithms and BC techniques. Section~\ref{sec:section4:solutions-for-iot} presents the latest solutions to IoT security and privacy threats, whereas research challenges for techniques based on ML and BC to solve security and privacy issues are presented in Section~\ref{sec:section5:challenges}. Finally, in Section~\ref{sec:section6:conclusion}, we conclude by presenting the gaps with some future directions. 
	\vspace{-0.2cm}
	\section{Threats in IoT}
	\label{sec:section2:threats-in-iot}
	IoT refers to a large number of heterogeneous sensing devices communicating with each other, either in a LAN or over the Internet \cite{Hussain2020_notSurvey}. IoT threats are different from conventional networks, significantly due to the available resources of end devices \cite{Qi2014}. IoT devices have limited memory and computational power, whereas the conventional Internet comprises powerful servers and computers with plentiful resources. Due to this, a traditional network can be secured by multi-factor security layers and complex protocols, which is what a real-time IoT system cannot afford. In contrast to traditional networks, IoT devices use less secure wireless communication media such as LoRa, ZigBee, 802.15.4, and 802.11a/b/n/g/p. Lastly, due to application-specific functionality and lack of common OS, IoT devices have different data contents and formats, making it challenging to develop a standard security protocol \cite{Makhdoom2018}. All these limitations make IoT prone to multiple security and privacy threats, thus opening venues for various types of attacks. 
	
	The probability of an attack in a network increases with the network size. Therefore, the IoT network has more vulnerabilities than a traditional network, for example, a company office. Additionally, IoT devices communicating with each other are usually multi-vendor devices with different standards and protocols. The communication between such devices is a challenge, which requires a trusted third party to act as a bridge \cite{Brass2018}. Moreover, several studies have raised the concern of regular software updates to billions of smart devices \cite{Fernandez-Carames2018, Lee2017_Sec}. 
	
	The computational resources of an IoT device are limited, so the capabilities of dealing with advanced threats are degraded. To summarize, IoT vulnerabilities can be categorized as \textit{specific} and \textit{common}. For example, vulnerabilities like \textit{battery-drainage attack}, \textit{standardization}, and \textit{lack of trust} are \textit{specific} to IoT devices, and Internet-inherited vulnerabilities can be regarded as \textit{common vulnerabilities}. Several IoT threats and their categorization have been introduced in the past \cite{Xiao2018, Restuccia2018, Mishra2018, Xiao2018c, butun2020IEEEsurvey}.  We discuss the most common threats in IoT reported in the past decade and attempt to classify them into security and privacy categories. 
	\vspace{-0.2cm}
	\subsection{Security Threats}
	
\begin{figure*}[t]
		\centering
		\begin{center}
			\vspace{-0.2cm}
			\includegraphics[width=4.0in,height=1.85in]{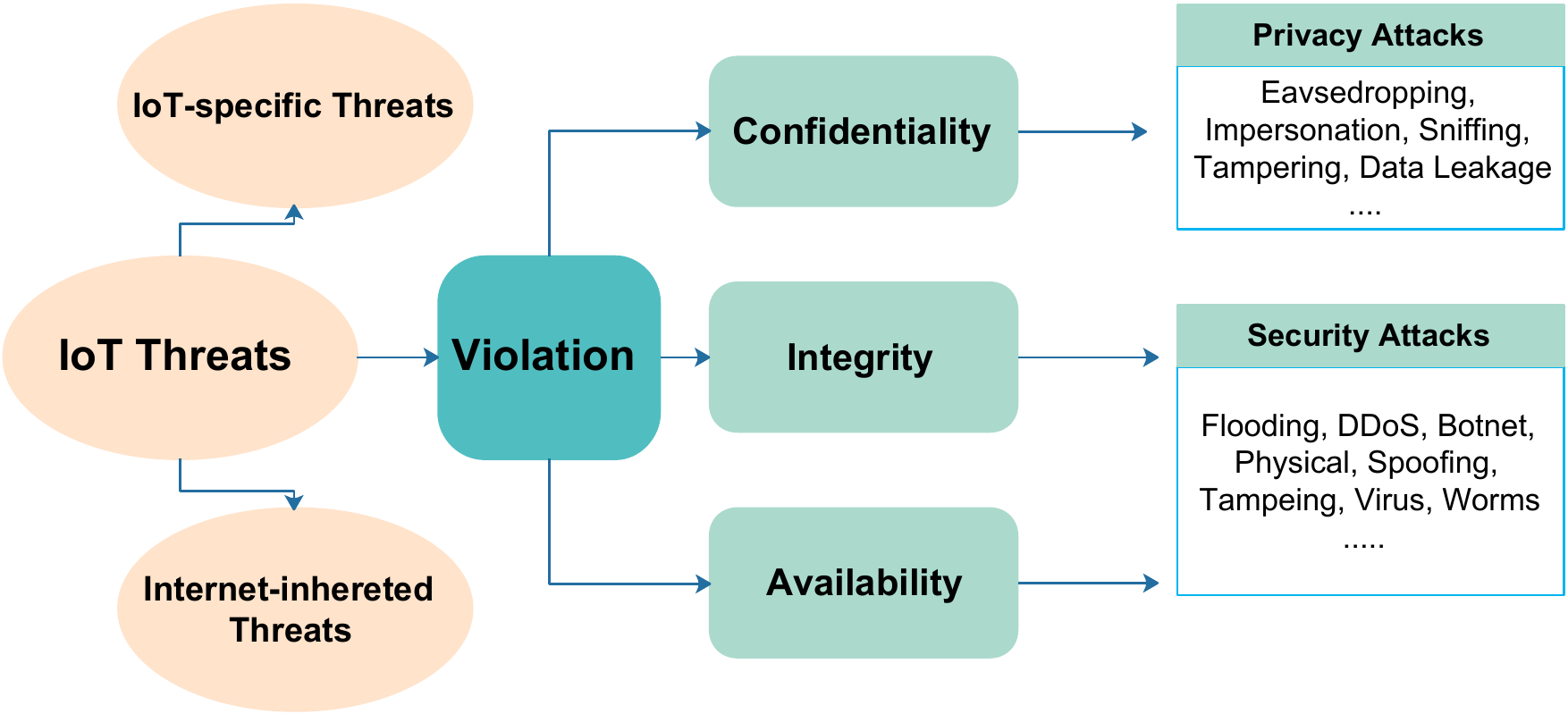}
			\vspace{-0.2cm}
			\caption{
				{Types of IoT threats may violate either of CIA triad; \textit{integrity} \& \textit{availability} are associated with security attacks, while \textit{confidentiality} compromise is known as privacy attack.}
			}
			\label{fig:TypesOfSecurityThreats}
		\end{center}
		\vspace{-0.4cm}
	\end{figure*}

	The fundamental concepts of security and privacy revolve around the CIA triad of \textbf{C}onfidentiality of the data, \textbf{I}ntegrity of data, and \textbf{A}vailability of the network \cite{Brewczynska2019, Yuen2019, Pavithra2019}. 
	In IoT, data can be anything, for example, a user's identity information, packets sent from a surveillance camera to a destination server, a command given by a user to its car using a key-fob, or a multimedia conversation between two people. Any unauthorized disclosure of data may result in a violation of either confidentiality, integrity, or availability. If a threat is impacting confidentiality, it is a privacy threat. The security threats affect both data integrity and network availability. Figure \ref{fig:TypesOfSecurityThreats} depicts different classes of security and privacy threats in IoT domains.  
	
	\vspace{-0.2cm}
	\subsubsection{Denial of Service}
	Denial of Service (DoS) has the most straightforward implementation among all the security attacks comparatively. Furthermore, the ever-growing number of IoT devices with weak security features has made DoS a favorite tool for attackers. The core objective of a DoS attack is to ingest the network with invalid requests, resulting in exhausting network resources, such as bandwidth consumption. As a result, the services are unavailable to genuine users. Distributed DoS (DDoS) is an advanced version of the DoS attack, where multiple sources attack a single target making it more difficult to trace and avoid the attack \cite{Diro2018, Abeshu2018, Tan2014, Tan2015, Sharma2017, Tselios2017}. Although there are different types of DDoS attacks, they all have the same objective. Few variants of DDoS attacks are SYN flooding \cite{Jing_DDoS_2019} (in which an attacker sends successions of SYN requests to a target), Internet Control Message Protocol (ICMP) attacks \cite{Elejla_DDoS_2019} (in which large number of ICMP packets are broadcasted using the victim's spoofed IP), crossfire attacks \cite{Rezazad_DDoS_2019} (using a complex and massively large-scale botnet for
	attack execution) and User Datagram Protocol (UDP) flooding attacks \cite{Kasinathan_DDoS_2013} (sending a large number of UDP packets to random ports on a remote victim). Botnet attack \cite{Moustafa2018} is a type of DDoS attack in an IoT network. The botnet is a network of IoT nodes (devices) that are compromised to launch an attack on a specific target, for example, a bank server. Botnet attack can be executed on different protocols, particularly Message Queuing Telemetry Transport (MQTT), Domain Name Server (DNS), and Hypertext Transfer Protocol (HTTP), as briefed in \cite{Moustafa2018}.
	Several techniques to prevent DoS in the IoT environment are suggested. Diro et al. \cite{Diro2018} have utilized the self-learning characteristics of Deep Learning (DL) methods to detect an attack in the fog-to-things environment. In another study, Abeshu et al. \cite{Abeshu2018} suggested controlling the DDoS attack by employing distributed DL on fog computing. Intrusion Detection System (IDS) by Tan et al. in \cite{Tan2014, Tan2015} is a series of research efforts to mitigate DDoS attacks using modern ML and DL algorithms. Sharma et al. and Tselois et al. \cite{Sharma2017, Tselios2017}, respectively, pointed out the issues of flooding in Software Defined Networks (SDN). The study highlighted that the SDN's top layer was prone to brute force attacks due to the lack of authentication in the plain-text TCP channel.
	\vspace{-0.2cm}
	\subsubsection{Man-in-the-middle}
	Man-in-the-middle (MiTM) attacks are one of the oldest attacks in the cyber world \cite{Swinhoe2019}. Spoofing and impersonation can be categorized as MiTM attacks. For example, a node \textit{X} intending to communicate with destination \textit{B} might be communicating with the MiTM attacker, who is impersonating to be destination B. Similarly, in SSL striping, an attacker can capitalize on such attacks to connect themselves with the server using an HTTPS connection, but with the target on an unsecured HTTP connection. Recently, many studies have focused on improving the security against MiTM attacks \cite{Ahmad2018, Aminanto2017, Chatterjee2018, Wang2017, ns_usenixsec2020}. Ahmad et al. \cite{Ahmad2018} discussed a healthcare scenario, where a patient gets an insulin dosage automatically. Such an application is vulnerable to MiTM attack, which can prove fatal. For instance, Tang et al.~\cite{ns_usenixsec2020} identified vulnerabilities in mobile apps' network service libraries, which can potentially expose apps' traffic to MiTM attacks. 
	Similarly, in line with the impersonation attacks, Chatterjee et al., \cite{Chatterjee2018} highlighted existing methods of authentication in wireless mobile devices that used a secret key. This key was saved in non-volatile memory and used for digital signatures or hash-based encryption. Apart from being vulnerable, this technique was power inefficient. Similarly, the most recent and widely used IoT protocol, called OAuth 2.0, suffers from cross-site-recovery-forgery (CSRF) attacks. The OAuth protocol manually authenticates devices, which is a time-consuming process. Another study by Wang et al. \cite{Wang2017} mentioned physical-layer security vulnerability in wireless authentication. They argued that the existing hypothesis test to compare radio channel information with channel record of \textit{Alice} to detect a spoofer \textit{Eve} in wireless networks, is often unavailable, particularly in dynamic networks. 
	
	\vspace{-0.2cm}
	\subsubsection{Malware} 
	{\textit{Malware}} is an abbreviation of \textbf{mal}icious soft\textbf{ware}. 
	Over the last couple of years, the number of IoT devices is growing along with frequent IoT software patches, which may be leveraged by an attacker to install malware on a device and perform malicious activities. Malware is generally understood to exist as a virus, spyware, worm, trojan horse, rootkit, or malvertising \cite{Makhdoom2018, zhao2019decade}. Smart home products, healthcare devices, and vehicular sensors are a few examples that can be compromised. Azmoodeh et al. \cite{Azmoodeh2018} studied malware on the Internet of Battlefield Things (IoBT). Such attackers are usually state-sponsored, better-resourced, and professionally-trained. Aonzo et al. \cite{Aonzo2017}, Feng et al. \cite{Feng2018}, and Wei et al. \cite{Wei2017} attempted to defend resource-constrained android devices from malware attacks by using different supervised ML algorithms. Studies in \cite{Lee2017_Sec, Sharmeen2018, Gu2018} provided a detailed analysis of malware detection and highlighted several security loopholes in the Android platform, especially on the application layer, which has applications with several types of components. 
	\vspace{-0.2cm}
	\subsection{Privacy Threats}
	
	{In addition to security threats, IoT users and their data are prone to privacy attacks, such as sniffing, de-anonymization, and inference attacks. In any case, the impact is on the confidentiality of data, where data can be at rest or in motion. In this section, we discuss various privacy attacks.}
	\vspace{-0.2cm}
	\subsubsection{MiTM}
	{We believe that MiTM attacks can be classified into Active MiTM Attacks (AMA) and Passive MiTM Attacks (PMA). The PMA passively listens to data transfer between two devices. Although the PMA violate privacy, they do not alter the data. An attacker with access to a device can silently observe for months before attempting the attack. With the growing number of cameras in IoT devices like toys, smartphones, and wristwatches, the impact of PMA, for example, eavesdropping and sniffing, is immense.  On the other hand, the AMA are actively involved in abusing the data acquired by either interacting with a user pretending to be someone else, for example, impersonation, or accessing a profile without consent, for example, authorization attack.} 
	\vspace{-0.2cm}
	\subsubsection{Data Privacy}
	
	{Similar to MiTM attacks, the data privacy attacks can be classified into \textit{Active Data Privacy Attacks} (ADPA) and \textit{Passive Data Privacy Attacks} (PDPA). Data privacy is related to data leakage {\cite{Wang2017_Privacy}}, data tampering, identity theft, and re-identification {\cite{Al-Rubaie2018}}. The re-identification attacks are also known as inference attacks and are based on de-anonymization attacks, location detection, and aggregation of information {\cite{Al-Rubaie2018}}. In these attacks, hackers' main goal is to gather data from multiple sources and reveal the targets' identities. Some attackers may use the collected data to impersonate an individual target {\cite{Xiao2018c}}. Any attack that alters data, such as data tampering, can be classified as ADPA, while the re-identification and data leakage are examples of PDPA.}     
	
	A comparison between various security and privacy threats, their types, their impacts, and possible solutions are summarized in Tables \ref{tab:SecurityThreatsTable} - \ref{tab:PrivacyThreatsTable}.
	
	\vspace{-0.4cm}
	\begin{table}[hb]
		\centering
		\caption{Security threats in IoT}
		\vspace{-0.2cm}
		\resizebox{\textwidth}{!}{\begin{tabular}{p{1.2cm} p{2.0cm} p{2.5cm} p{3.5cm} p{2.5cm} p{4.0cm}} \toprule
				{\bf Threat} & {\bf Impact} & {\bf Attack} & {\bf Type} & {\bf Layer of Impact} & {\bf Solution} \\ \midrule
				Security & Availability & DoS & Flooding & Physical, MAC & Multiple \\
				&  &  & DDoS & Physical, MAC & Multiple \\
				&  &  & Botnet & Physical, MAC & Multiple \\	
				&  & Physical & Damage & Physical & Physical Security \\
				&  &  & Environmental & Physical & Shielding \\
				&  &  & Power Loss & Physical & uninterrupted power \\
				&  &  & Hardware Failure & Physical & Backup \\
				&  &  & Tampering & Physical & Physical Security \\
				& Integrity & MiTM & Sybil Attack & Physical, MAC, Network & code attestation, radio resources testing, key pool \\
				& & &Spoofing&Network&anti-spoofing software\\
				&&&message tamper&&\\
				&&Malware&Injection&Application&\\
				&&&Virus&Application&\\
				&&&Worms&Application&\\
				\bottomrule
		\end{tabular}}
		\vspace{-0.4cm}
		\label{tab:SecurityThreatsTable}
	\end{table}
	
	%%%%%% Table of Privacy Threats in IoT
	\vspace{-0.4cm}
	\begin{table}[ht]
		\centering
		\caption{Privacy threats in IoT}
		\vspace{-0.2cm}
		\resizebox{\textwidth}{!}{\begin{tabular}{p{1.2cm} p{2.0cm} p{2.5cm} p{3.5cm} p{2.5cm} p{4.0cm}} \toprule
				{\bf Threat} & {\bf Impact} & {\bf Attack} & {\bf Type} & {\bf Layer of Impact} & {\bf Solution} \\ \midrule
				Privacy & Confidentiality & MiTM & Eavesdropping & Network & Encryption \\
				&  &  & Impersonation & Network &  Encryption \\
				&  &  & Sniffing & Network & Encryption \\	
				&  &  & Authroization & Application & Access Control \\
				& & Data Privacy & Data Leakage & Multiple & \\ & & & Re-identification & Multiple & data suppresion, generalization, noise addition\\
				&&&Data tampering&Multiple&anonymization\\
				&&&Identity Theft&Multiple&anonymization\\
				&&Others&Poodle&Transport&Use TLSv1.2\\
				&&&Heartbleed&Transport\\
				&&&Freak&Transport&Turnoff export ciphersuit options in browser\\
				\bottomrule
		\end{tabular}}
		\vspace{-0.2cm}
		\label{tab:PrivacyThreatsTable}
	\end{table}
	\vspace{-0.4cm}
	\section{Literature Survey}
	\label{sec:section3:lit-survey}
	This Section provides an existing literature review and categorizes the efforts done based on ML algorithms and BC techniques to address IoT security and privacy issues. This Section is divided into two subsections, i.e., ML algorithms and BC techniques. 
	
	\vspace{-0.2cm}
	\subsection{{Existing review papers using Machine Learning Algorithms as a solution}}
	
	Hackers are getting sophisticated with the evolving technology, making traditional methods of attack-prevention cumbersome. The defense becomes more challenging for a resource-constraint IoT device. {To help in detecting these attacks, one of the widely used tools is ML algorithms. ML can be defined as the ability to deduce knowledge from data, and adjust the output of an ML model, based on that acquired knowledge \cite{Hussain2020_notSurvey}. ML makes machines smart enough by learning from their past results and refining them to achieve improved results \cite{Haque2020}}. Several ML algorithms have proven extremely helpful in mitigating security as well as privacy attacks. In the following subsections, we discuss these approaches in detail.
	
	\subsubsection{Security Efforts}
	
	The technology has improved data communication and networking techniques over the Internet. We have now state-of-the-art software-based configurable devices called Software Defined Network (SDN) that can be customized to meet a customer's needs. In this scenario, Restuccia et al. \cite{Restuccia2018} attempted to present the taxonomy of existing IoT security threats and their solutions in SDN  using the ML algorithms. They also suggested that since the main task of an IoT system is to collect data from IoT devices, it is feasible to divide the data collection process into three steps, namely IoT authentication, IoT wireless networking, and IoT data aggregation \& validation. The study gave a brief review of ML algorithms used to mitigate the security attacks, e.g., to detect cross-layer malicious attacks, Bayesian learning is used, and to assess the validity of data, neural networks are used. However, the study lacks an in-depth analysis of the rest of the ML algorithms.
	
	Sharmeen et al. \cite{Sharmeen2018} aimed to assist application developers in using Application Program Interfaces (APIs) safely, during the development of applications for Industrial IoT networks. {To detect malware, the authors suggested that the ML model could be trained by using three types of features including static, dynamic, and hybrid}. A detailed analysis of each feature type is done using performance metrics of a dataset, features extraction technique, features selection criteria, accuracy, and detection method. Several detection methods for each feature set were analyzed, but the commonly used were RF, SVM, KNN, J48, and NB. Sharmeen et al. \cite{Sharmeen2018} concluded that hybrid analysis offered flexibility in choosing both the static and dynamic features to improve accuracy in the detection process. However, this paper is limited to one application (android device) and one security threat (malware).
	
	{Costa et al. \cite{Costa2019} selected papers between 2015 to 2018 and claimed that no work has presented an in-depth view of the application of ML in the context of IoT intrusion detection. The study reviewed the latest as well as traditional ML-based algorithms to improve IoT security. They also presented the most commonly used datasets and methodologies employed in the paper related to IoT security. The paper however has not reviewed the latest IoT security or privacy threats. }
	
	{Similarly, Chaabouni et al. \cite{Chaabouni2019} also focused on the IoT based network intrusion detection systems. The authors presented IoT architecture and layer-wise attacks, and classified them by layers (\textit{perception layer}, \textit{network layer} and \textit{application layer})  as well as design challenges (such as \textit{heterogeneity, mobility, trust and privacy, resource constraints, connectivity and data interchange}). The traditional mechanisms to protect IoT were described, and the study focused on Anomaly and Hybrid Network IDS (ANIDS) for IoT systems. A detailed comparison of traditional NIDS for IoT systems architecture, detection methodologies, and experimental results was provided. The study further presented how the Learning-based NIDS for IoT could overcome the challenges faced by their equivalent traditional IoT systems. Finally, top IoT NIDS proposals were compared with a focus on ML algorithms. }
	
	All of the above papers, as depicted in Table \ref{table:ExistingSurveysMLSecurity}{,} are limited to security threats with a focus on ML as a tool in solving the security issues. {Our paper, as depicted in Figure \ref{fig:2in1}a, covers a broader scope - addressing security and privacy issues in IoT domains using ML and BC.}
	
	%Existing Surveys on IoT Security Using MLA
	\vspace{-0.2cm}
	\begin{table}[!ht]
		\centering
		\caption{{List of survey papers on IoT security leveraging Machine Learning Algorithms.}}\vspace{-0.2cm}
		\resizebox{\textwidth}{!}{\begin{tabular}{|l|l|l|l|}
				\hline
				
				\textbf{Ref.} & \textbf{Security Threats} & \textbf{{Proposed } solution(s)} \\ \hline
				
				{Restuccia et al.} \cite{Restuccia2018} & DoS, MiTM & A taxonomy and survey of IoT security research and their ML-based solutions \\ \hline
				
				Sharmeen et al. \cite{Sharmeen2018}  & Malware  &  Analysis of malware detection for Android mobile  \\ \hline	
				
				{Chaabouni et al. \cite{Chaabouni2019}}   & Multiple &  A detailed analysis of traditional and ML-based NIDS for IoT. \\ \hline
				
				{Costa et al. \cite{Costa2019}}   & Multiple &  In-depth review of ML applications in the context of IoT intrusion detection. \\ \hline
				
		\end{tabular}}\vspace{-0.4cm}
		
		\label{table:ExistingSurveysMLSecurity} \centering
	\end{table}

	\subsubsection{Privacy Efforts}
	
	Machine Learning extracts useful information from the raw data, while privacy is preserved by concealing the information \cite{Ji2014}. According to Al-Rubaie et al. \cite{Al-Rubaie2018}, ML system has three modules: {(i)} input, {(ii)} computation, and {(iii)} output. The study further claimed that privacy could only be preserved if all three modules were under the ownership of a single entity. Nowadays, the data is collected {worldwide by billions of IoT devices such as smart-phones, health monitoring sensors, speed cameras, and temperature sensors, hence a \textit{single-ownership} condition cannot be maintained}. This issue spurred interest in researchers to work towards proposing newer and improved privacy-preserving ML algorithms.
	{For instance, the lack of privacy protection mechanisms in a VANET environment was raised by Zhang et al. \cite{Zhang2018}. In VANET, Vehicle nodes tend to learn collaboratively, raising privacy concerns, where a malicious node can obtain sensitive data by inferring from the observed data. A single node has limited computational and memory resources. The solution was presented by using collaborative IDS with distributed ML algorithms and resolving the privacy issues by proposing the concepts of dynamic differential privacy to protect the privacy of a training dataset.}
	
	{People traffic monitoring systems and healthcare services are two of the most common IoT sensing technologies}, which need continuous improvements. The most effective and useful data for such applications is {directly collected from the users through} Mobile CrowdSensing (MCS). Xiao et al. \cite{Xiao2018c} reviewed the privacy threats involved in MCS, where the information of interest is extracted, and the participants upload sensing reports of their surroundings to the MCS server. This information-sharing poses significant privacy threats to the participants and the MCS server. The system is prone to privacy leakage (which is related to user’s personal information), faked sensing attacks (sending fake reports to the server to reduce the sensing efforts) and advanced persistent threats (causing privacy leakage over an extended period). The survey suggested Deep Neural Network (DNN) and Convolutional Neural Network (CNN) for privacy protection, and Deep Belief Network (DBN) and Deep Q-Network (DQN) for counter-measuring faked sensing. However, the review was limited to only one application (MCS).    
	
	\vspace{-0.2cm}
	\subsection{{Existing review papers using Blockchain as a solution}}
	
	Blockchain, often confused by some as a synonym to bitcoin, is the technology behind this infamous crypto-currency. It is a distributed ledger which stores the data in blocks. These blocks are in order and linked with each other cryptographically forming a chain in a way that makes it computationally infeasible to alter the data in a particular block \cite{Chapron2017}. This mechanism ensures immutation, decentralization, fault-tolerance, transparency, verifiability, audit-ability, and trust \cite{Christidis2016, Panarello2018}. {There is no single consensus on the types of BC but most commonly they are public, private, and consortium. Public or permission-less BC is open to everyone, so anyone can access them \cite{Fernandez-Carames2018}.  On the other hand, Private/permissioned blockchains are controlled by one or few, hence not everyone can access them, the transactions here are faster and only the selected few are authorized to approve a transaction, hence reaching a consensus. Several reviews and survey papers \cite{Banerjee2018, Khan2018, Reyna2018, Meng2018, Panarello2018, Kshetri2017, Christidis2016, Makhdoom2018, Tschorsch2015, Dunphy2018, Fernandez-Carames2018, Yu2018, Li_BCsurvey_2017} are published to highlight the importance of the BC techniques and could be a good source for those who are interested to read more about BC in detail.}  A detailed comparison of current work is shown in Table \ref{table:ExistingSurveysBCSecurity} and \ref{table:ExistingSurveysBCPrivacy}. Most of these works discussed either security or privacy issues. In this Section, we present the current literature reviews on achieving security and privacy in IoT using BC techniques as a tool.
	
	%Existing Survey on IoT Security Using Blockchain        
	\begin{table}[]
		\centering
		\vspace{-0.2cm}
		\caption{{List of survey papers on IoT security leveraging Blockchain techniques.}}\vspace{-0.2cm}
		\resizebox{\textwidth}{!}{%
			\begin{tabular}{|l|l|l|l|}
				\hline
				\textbf{Ref.} & {\bf Year}&  \textbf{Security Threats} &  \textbf{Comments} \\ \hline
				Banerjee et al. \cite{Banerjee2018}& 2018& Several   &   {Classified post-2016 literature \& discusses BC-based solutions} \\ \hline
				
				Khan et al. \cite{Khan2018}& 2018 &  {Key management} &  Categorization of threats \& their BC-based solution{s} \\ & & & access control were presented
				\\ \hline
				
				Reyna et al. \cite{Reyna2018}& 2018  &  DoS  &  Challenges \& Analysis of BC in IoT devices were mentioned   \\ \hline
				
				{Panarello et al. \cite{Panarello2018}} & {2018} & {Multiple}   & Comprehensive BC-IoT integrated security challenges and \\ & & & emerging solutions were discussed  \\ \hline
				
				Kumar et al. \cite{Kumar2018} &2018&  MiTM & How BC can be a solution for IoT security issues, is discussed  \\ \hline
				
				{Kouicem et al. \cite{Kouicem2018}} & {2018} &  {Multiple} & {Provided BC-based solution to attain a ``trio'' of \textit{anonymity}}, \\ & & & {\textit{unlinkability}, and \textit{intractability}}    
				\\ \hline
				
				{Wang et al. \cite{Wang2019}} &2019&  Multiple & IoT layer-wise attacks discussed. BC-based security solutions \\ & & & for IoT applications were discussed   \\ \hline
				
				{Ali et al. \cite{Ali2019}} & 2019 &  DoS & Reviewed latest proposed BC-based IoT security solutions   \\ \hline

				{Hassija et al. \cite{Hassija2019}}  & 2019  & Multiple &  A detailed survey of existing IoT security solutions is presented \\ \hline
				
		\end{tabular}}
		\vspace{-0.6cm}
		\label{table:ExistingSurveysBCSecurity}
	\end{table}
	
	\subsubsection{Security Efforts}

	Security has been the prime focus of attention for any IoT use cases. Lots of work based on BC techniques have emerged to solve security issues in the IoT domain. A study on IoT security was presented by Banerjee et al. \cite{Banerjee2018}, which is classified into security techniques such as intrusion detection and prevention system (IDPS), collaborative security, and predictive security. Furthermore, IDPS are classified by approaches, network structure, and applications. After that, collaborative security and predictive security are discussed in detail. In the same study, collaborative security techniques are classified by network structures and applications. Sequel to this study, the integrity of existing IoT datasets is highlighted, and the authors suggested that a BC-based standard should be developed to ensure integrity in the shared datasets. 
	
	In another study by Khan et al. \cite{Khan2018}, security issues related to key management, access control, and trust management in IoT are discussed. Khan et al. \cite{Khan2018} categorized the security threats into IoT layers and presented their BC-based solutions. The IoT security issues were classified as \textit{low-level}, \textit{intermediate-level}, and \textit{high-level} security issues. Khan et al. \cite{Khan2018} believes that jamming adversaries, insecure initialization, spoofing, vulnerable physical interface, and sleep deprivation attacks are the low-level security issues. Whereas, replay, RPL routing attacks, sinkhole, Sybil attack on intermediate layers, transport-level end-to-end security, session establishment, and authentication are intermediate-level security issues. The high-level security issues are insecure interfaces, CoAP security with {I}nternet, vulnerable software, and middleware security. The study then provided a comprehensive mapping of all the above problems with the affected layers of IoT architecture and proposed solutions for each one of them. In the end, the authors discussed how BC techniques could be used to address and solve some of the most pertaining IoT security problems. This survey highlighted the security risks involved in each IoT layer but lacked the discussion of providing BC-based solutions for these security threats. 
	
	Similarly, Reyna et al. in \cite{Reyna2018} analyzed how BC techniques could potentially improve the security (data reliability) in the IoT. The study mentioned security threats as one of the challenges for BC techniques. The security threats mentioned in the study were majority attacks, double-spend attacks, and DoS attacks. The study also provided {highlights} about the integration of IoT with BC techniques, BC applications and BC platforms.  
	However, the study did not cover several other security attacks related to IoT, which was a limitation of this survey.
	
	{On the other hand Panarello et al. \cite{Panarello2018} comprehensively reviewed BC consensus protocols in addition to security challenges and recent developments in IoT and BC integration. The past literature was categorized based on application areas, which were supported by an extensive survey of the latest BC-based solutions.}
	
	{Kumar et al. \cite{Kumar2018} presented a brief overview of issues and challenges in IoT security, such as spoofing and false authentication. Some of the advantages of BC for large scale IoT systems are tamper-proof data, trusted and reliable communication, robustness, and distributed and delegated data sharing. Sequel to that, the study has discussed the application-wise BC-based IoT challenges.}
	
	{The authors of \cite{Kouicem2018} highlighted security issues in IoT and provided their BC-based solutions. The study first highlighted the IoT security requirements and its challenges in six different application domains, like smart cities, healthcare, smart grids, transport, smart homes, and manufacturing. The authors comprehensively discussed the taxonomy of IoT security solutions such as confidentiality and availability. They also investigated the analysis of techniques that were suitable for each IoT application. }
	
	{Wang et al \cite{Wang2019} highlighted the limitation of IoT security and provided comprehensive security analysis on end devices, communication channels, network protocols, sensory data, DoS attack, and software attacks. After presenting the existing BC technologies, the application of BC for IoT and their challenges were discussed. The study also briefly discussed the security of IoT applications using BC.}
	
	{The potential benefits and motivations for developing a BC-based IoT framework are \textit{resilience, adaptability, fault tolerance, security and privacy, trust} and \textit{reduced maintenance cost}. \cite{Ali2019}
		The study mentioned that the centralized IoT model is prone to DDoS attacks. Moreover, due to its architecture, it has a single point of failure which is a threat to the availability of IoT services. Current IoT security solutions are centralized because they involve trusting in third party security services which bring in data integrity issues. Ali et al. highlighted how all of these issues can be solved by using BC-based IoT security solutions. }
	
	{A comparison of IT and IoT security, followed by a comprehensive classification of IoT applications and their security and privacy issues were discussed by Hassija et al. \cite{Hassija2019}. The study even went ahead and discussed various possible security threats in IoT applications for four layers, i.e., (i) sensing, (ii) network (iii) middle-ware, and (iv) application.  In the recommendations to improve the IoT security, BC was also mentioned as one of the solutions. This paper is probably the closest to our work, however, it is security-biased and does not focus on the IoT privacy issues in detail.  }
	
	%Existing Survey on IoT Privacy Using Blockchain        
	\begin{table}[]
		\centering
		\caption{List of survey papers on IoT privacy leveraging Blockchain techniques.}\vspace{-0.2cm}
		\resizebox{\textwidth}{!}{\begin{tabular}{|l|l|l|l|l|}
				\hline
				\textbf{Ref.} &{\bf Year}&  \textbf{Privacy Threats}&  \textbf{Comments} \\ \hline
				Kshetri et al. \cite{Kshetri2017} & 2017   &  {Identity} management  & {Highlighted how BC is superior to the current IoT ecosystem} \\ \hline

				Kumar et al. \cite{Kumar2018} &2018& Spoofing, authentication & Presented IoT security and privacy issues and how BC can be a solution.  \\ \hline
				
				{Kouicem et al. \cite{Kouicem2018}} & {2018} & {Data Privacy} & {Provided BC-based solution to attain a "trio" of \textit{anonymity},} \\ & & & {\textit{unlinkability}, and \textit{intractability}}    
				\\ \hline
				
				{Zhu et al. \cite{Zhu2018}} & {2019} &  Data Privacy & Highlighted challenges in traditional IdM systems and reviewed \\ &&& their BC-based solutions   
				\\ \hline
				
				{Hassan et al. \cite{Hassan2019}} & {2019} &  Multiple & Comprehensively surveyed privacy preservation techniques of \\ &&& BC-based IoT systems from application and implementation  
				\\ \hline
				
				{Ali et al. \cite{Ali2019}} &2019&  Data Privacy, MiTM & Reviewed latest proposed BC-based IoT privacy solutions   \\ \hline
				
		\end{tabular}}
		\vspace{-0.2cm}
		\label{table:ExistingSurveysBCPrivacy}
	\end{table}
	
	\raggedbottom
	
	A table of the existing surveys focused on security using BC techniques for IoT applications is compiled in Table \ref{table:ExistingSurveysBCSecurity}.
	
	\subsubsection{Privacy Efforts}
	
	{Previous studies, such as \cite{Ferretti2019}, used strong cryptographic measures to protect against malicious third parties and provided accountable access to IoT. However, they did not use either ML or BC as one of their tools. Kshetri et al. \cite{Kshetri2017} highlighted how BC techniques can offer better privacy-preserving solutions as compared to a traditional network for cloud-based services. It also highlighted the superiority of BC in identity management and the provision of access control. The study demonstrated how an attack on the IoT network could be contained using BC techniques. However, a comprehensive privacy-preserving IoT threat model using BC techniques was missing in this literature.} 
	
	{ Kumar et al. \cite{Kumar2018} presented a brief overview of issues and challenges in IoT privacy, such as data sharing. Sequel to that the work suggested the BC-based solutions to these challenges and discussed several application areas for BC implementation. Although the work discussed challenges to BC in the IoT application, it lacks a comprehensive discussion on the latest IoT security and privacy threats.}%. was missing.}
	
	{The authors of \cite{Kouicem2018} highlighted privacy issues in IoT and provided their BC-based solutions. The main goal of privacy-preserving techniques was to attain a ``trio'' of \textit{anonymity}, \textit{unlinkability}, and \textit{intractability}. The main security services, for example, \textit{confidentiality}, \textit{privacy} and \textit{availability}, were addressed based on traditional cryptographic approaches. The study addressed the issues of data-sharing, data privacy, and user’s behavior in IoT, and discussed their solutions, for example, data tagging, zero-knowledge proof, pseudonyms, and k-anonymity model. }
	
	{Zhu et al \cite{Zhu2018} highlighted privacy vulnerabilities in a traditional Identity Management (IdM) system, especially due to their centralized architecture, and reliability on the so-called trusted third parties. These vulnerabilities may result in several privacy attacks such as phishing and data leakage. The authors argued that traditional IdM systems can not be directly transplanted to IoT environments due to some native IoT characteristics such as scalability, mobility, and compatibility. Sequel to that the study highlighted the privacy challenges in traditional IdM systems and reviewed their BC-based solutions.}
	
	{Hassan et al. \cite{Hassan2019} provided a detailed overview of privacy issues in BC-based IoT systems. The privacy attacks related to BC-based IoT networks such as \textit{Address reuse, Deanonymization, Sybil attack, Message Spoofing}, and \textit{Linking attacks} were highlighted. The work also discussed the implementation of the five most popular privacy preservation strategies (\textit{Encryption, Smart Contract, Anonymization, Mixing,} and \textit{Differential Privacy}) within BC-based applications.}
	
	{Ali et al. \cite{Ali2019} reviewed the IoT privacy issues and their latest BC-based solutions. They raised the privacy concerns in a centralized IoT model such as Data privacy and data confidentiality. The existing centralized privacy solutions such as \textit{using a privacy broker}, \textit{using group signatures}, applying \textit{k-anonymity}, and pseudonyms, were all heavily dependent on third parties for their services. To counter these issues, the study offered a comprehensive review of the BC-based IoT privacy solutions. }
	
	\vspace{-0.2cm}
	\section{Solutions to IoT Threats}
	\label{sec:section4:solutions-for-iot}
	Since the inception of the first virus (Creeper) in 1970 until the hack of Whatsapp on 15th May 2019 {\cite{Creeper, Policy2019} and later, security specialists have mitigated zero-day security or privacy threats \cite{ikram2019chain, Tan2014, zhao2019decade}}. 
	{ Regarding this, several solutions have been proposed to mitigate security and privacy issues. However, in this Section, we focus on the recent literature proposing secure and privacy-preserving techniques for the IoT domain. We discuss the solutions offered by first using ML algorithms as a tool,  {\it then} by utilizing BC techniques, and {\it finally} by the fusion of both.}
	
	\vspace{-0.2cm}
	\subsection{Existing Solutions Using Machine Learning Algorithms}
	
\begin{figure}[]
		\centering
		\begin{center}
			\vspace{-0.2cm}
			\includegraphics[width=3.5in,height=3.5in]{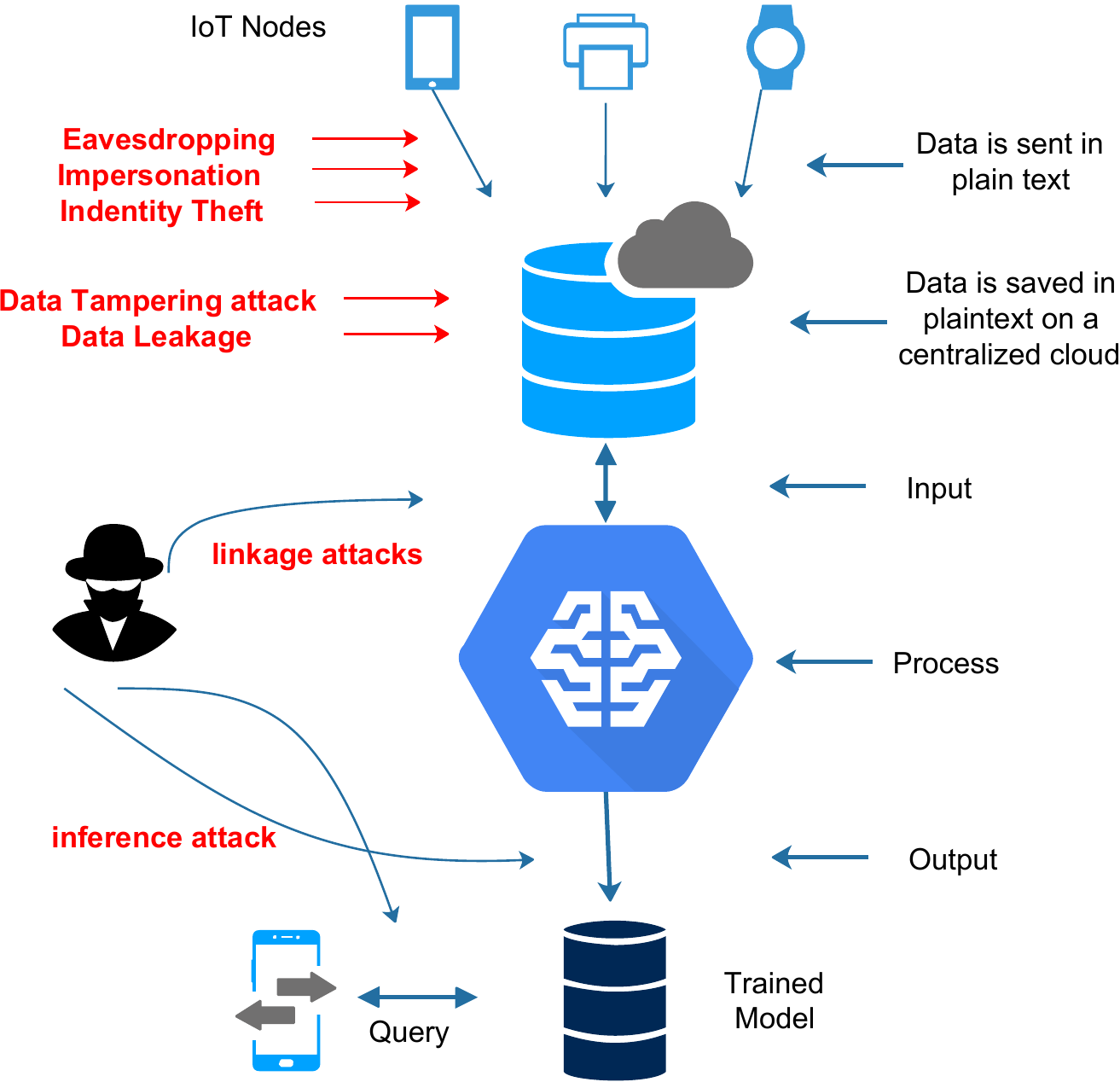}
			\vspace{-0.2cm}
			\caption{{An illustration of ML threat model for IoT: A ML model is prone to several attacks at either (i) \textit{input}, (ii) \textit{process}, or (iii) \textit{output} stages.}}
			\label{fig:TM4IoT}
		\end{center}
		\vspace{-0.6cm}
	\end{figure}

	ML is used as a data processing pipeline in any framework. For example, data traffic entering a network can be analyzed by an ML model to make an informed decision. The main components of the ML threat model for IoT are shown in Fig{ure} \ref{fig:TM4IoT}. Additionally, the figure gives an overview of target points, such as input and output, for an attacker. The input data from source to IoT nodes, and IoT nodes to ML model can experience exploratory or poisoning attacks. At the output, integrity and inversion attacks are possible \cite{LiuShui2018}. Therefore, for a whole system to be completely immune to attacks, it must be secured as well as privacy-preserved.
	
	\subsubsection{Security efforts}
	
	Several security solutions have been proposed using ML algorithms as a tool, as shown in Table \ref{tab:SecSolML}.
	To deal with the flooding attacks, Diro et al. \cite{Diro2018} argued that fog-computing reduced the risk of eavesdropping and MiTM attacks by restricting the communication to the proximity of IoT devices. Capitalizing on this idea, they used the Long Short Term Memory (LSTM) algorithm in their model as it can remember the older data. For binary classification, they compared their results with LR using ISCX2012 dataset, which had 440,991 normal traffic instances and 71,617 DoS attack instances. The DL model LSTM took considerably more time to train than LR, but its accuracy was 9\% better. The second dataset used was AWID from \cite{Kolias}, and consists of normal traffic instances (1,633,190 training and 530,785 tests), injection attack instances (65,379 training and 16,682 tests), flooding attack instances (94848 training and 8097 testings) and impersonation attack instances (48,522 training and 20,079 testings). After comparing LSTM against softmax for multi-class classification, the resultant accuracy obtained was 14\% improved. 
	
	In a similar study, Abeshu and Chilamkurti highlighted that the resource constraints of an IoT device made it a potential threat to DoS attacks \cite{Abeshu2018}. Classic ML algorithms are less accurate and less scalable for cyber-attack detection in a massively distributed network such as IoT. Such a massive amount of data produced by billions of IoT devices enable the DL models to learn better than the shallow algorithms. The authors of \cite{Abeshu2018} argued that most of the employed DL architectures had used pre-training for feature extraction, which could detect anomalies and thus reduced the workload of a network administrator. However, their work was focused on distributed DL through parameters and model exchange for the applications of fog computing. Fog computing reduced the load of computing power and storage space from the IoT devices. It is, therefore, the ideal spot where an intrusion can be detected. The existing Stochastic Gradient Descent (SGD) for fog-to-things computing needs parallel computing. Thus, the centralized SGD will choke due to the massive amount of data in IoT. Therefore the study proposed a distributed DL-driven IDS using NSL-KDD dataset, where the stacked auto-encoder (SAE) was used for feature extraction, and soft-max regression (SMR) was used for the classification. Their study proved that the SAE as a DL worked better than traditional shallow models in terms of accuracy (99.27\%), FAR and DR. Both Diro et al. \cite{Diro2018} and Abeshu et al. \cite{Abeshu2018} proved that the DL algorithms performed better than shallow ML models. 
	
	As a first attempt to DoS detection, Tan et al. \cite{Tan2014} used triangle-area-based technique to speed up the feature extraction in Multivariate Correlation Analysis (MCA). Features were generated to reduce the overhead,  using the data that entered the destination network. Along with this, the ``triangle area map'' module was applied to extract the geometrical correlations from a pair of two distinct features to increase the accuracy of zero-day attack detection. In an attempt to improve their results from \cite{Tan2014}, Tan et al. \cite{Tan2015} used Earth Mover's Distance (EMD) to find the dissimilarities between observed traffic and a pre-built normal profile. The network traffic was interpreted into images by feature extraction using MCA and analyzed to detect anomalies using KDDCup99 and ISCX datasets. Using the sample-wise correlation, the accuracy of their results obtained was 99.95\% (KDD) and 90.12\% (ISCX). However, the study  neither revealed the data size nor the effects of varying sample sizes. Moreover, MCA assumed the change to be linear, which was not a realistic approach. Another form of DoS attack in IoT is called a botnet attack, which was explained earlier in Section~\ref{sec:section2:threats-in-iot}. To prevent botnet attacks against HTTP, MQTT (Message Queuing Telemetry Transport), and DNS, the authors of \cite{Moustafa2018} developed an IDS, which is an ensemble of DT, NB, and ANN. Since the correntropy values of benign and malicious vectors were too close, it was decided to use DT, NB, and ANN as they could classify such vectors efficiently. The performance metrics were detection-rate and false-positive rate, for which their proposed ensemble was better than every individual algorithm in that ensemble. For the datasets of UNSW and NIMS, the accuracies achieved were 99.54\% and 98.29\%, respectively.     
	
	Similar to DoS attacks, the MiTM attacks are one of the most frequently occurring attacks in an IoT network. In regard to this, a lot of technical solutions have been proposed for several applications. The authors \cite{Ahmad2018} have used LSTM RNN to prevent the impersonation attacks in a smart healthcare scenario, since traditional feedforward neural networks cannot capture the sequence and time-series data, due to their causal property. Moreover, the researchers solved the vanishing gradient issue of RNN algorithm and improved accuracy. At first, the predicted value was calculated based on the dataset log of three months (for a patient who is taking insulin injections). If the predicted and calculated values differed for more than a certain threshold, then by using the combination of DL and gesture recognition, the correct dosage was ensured. However, the detail of the model and analysis was missing in their work. 
	
	Similarly in another scenario to prevent the impersonation attacks, the authors of \cite{Chatterjee2018} utilized Physical Unclonable Function (PUF), which is an inherent characteristic of silicon chips that is unique and can be used as a basis of authentication in RF communication. During the manufacturing phase, every transmitter inherits some unique features called \textit{offset} from an ideal value. The authors have used these offsets as their features to recognize the device, train their system on it, and then detect the accuracy. Using ANN MATLAB toolbox, the performance metrics were calculated. With the help of ML, the simulation results could detect 4,800 nodes transmitters with an accuracy of 99.9\% and 10,000 nodes under varying channel conditions, with an accuracy of 99\%. The proposed scheme can be used as a stand-alone security feature, or as a part of traditional multi-factor authentication. PUF is inherent and inexpensive and can significantly benefit IoT, wherein each wireless sensor's physical values can be stored in a secure server replacing traditional key-based authentication. However, the authors in their approach have assumed the server storing the PUF values is safe. Aminanto et al. used an unsupervised ensemble of ML algorithms using SVM, ANN, and C4.5 for feature extraction and ANN as the classifier \cite{Aminanto2017}. In their process of deep-feature extraction and selection (D-FES), first, they used SAE to extract the features, then SVM, ANN, and C4.5 were used for feature selection, and finally, ANN was used to classify. The study achieved an accuracy of 99.92\% by using AWID dataset,  on which an earlier study by Kolias et al. \cite{Kolias} had the worst accuracy for impersonation attack.  
	
	According to {\it Statista} \cite{EMarketer2016}, mobile phone users would reach close to three billion by 2020. This increase in usage made mobile phones vulnerable to the malware attack \cite{Azmoodeh2018, Aonzo2017, Wei2017, Feng2018, Sharmeen2018, Wang_DriodEnsemble2018}. Azmoodeh et al. \cite{Azmoodeh2018} believed that OpCodes could be used to differentiate benign-ware and malware. Class-wise Information Gain (CIG) is used for feature selection because the global feature selection causes imperfections, and even reduces system efficiency especially when the dataset is imbalanced. They also claimed that this combination of OpCode and DL for IoT had never been explored. Using Eigenspace and deep convolutional networks algorithms, 99.68\% accuracy was achieved, with precision and recall rates of 98.59\% and 98.37\%, respectively. Similarly, to mitigate malware, Wei et al. \cite{Wei2017} extracted the features using the dynamic analysis technique. They used application functional classification to train the classifier for clean and malicious data, while, in the testing phase, kNN was used to divide data into known categories. J48 decision tree and NB were used to perform 10-fold cross-validation. Depending on the performance metric, the study claimed 90\% accuracy. 
	
	Contrary to dynamic analysis\cite{Wei2017}, the authors of \cite{Aonzo2017} used static analysis techniques for feature extraction considering all the Application Platform Interfaces (API) that were not studied previously. Feature selection was made manually based on the most-used features by the previous researchers. They claimed the accuracy of 98.9\% with the second biggest malware testbed dataset ever used. As the intrusion techniques were getting sophisticated, the static analysis became invalid, and it was therefore required to use a dynamic scheme \cite{Feng2018}. With the static analysis techniques, the attackers adopted deformation technologies, which could bypass the detection while dynamic analysis methods were promising due to its resistance to code transformation techniques. The authors of \cite{Feng2018} proposed a new framework, called EnDroid, based on these issues. The proposed model used ``Chi-Square'' for feature extraction, five different algorithms (decision tree, linear SVM, extremely randomized trees, random forest \& boosted trees) as an ensemble for base-classification, while LR was used as meta-classifier. For the dataset, a combination of ``AbdroZoo'' and ``Drebin'' datasets was utilized so that an accuracy of 98.2\% was achieved. Wang et al. argued that most of the existing literature on malware detection was based on static string features, such as permissions and API usage extracted from the apps \cite{Wang_DriodEnsemble2018}. However, since malware had become sophisticated, using a single type of static feature might result in a false-negative. In their proposed model - DriodEnsemble,  a fusion of string and structural features was utilized to detect Android malware. Using an ensemble of SVM, kNN, and RF, the model was evaluated against 1,386 benign apps and 1,296 malapps. The study proved to have attained an accuracy of 98.4\%, which was better than detection accuracy (95.8\%) using only string features, while the accuracy obtained with only structural features was 90.68\%.
	
	\begin{sidewaystable*}
		
		\caption{Existing IoT security solutions using machine learning algorithms. }%Here, the notations in `Dataset' column represent: I: ISCX2012, A: AWID, N: NSL-KDD, K: KDDCUP99, U: UNSW-NB15, NB: NIMS botnet, P: Private, AWI: Aegan WiFi Intrusion, Ab: AbdroZoo, D: Drebin, C: CTU-13, and Ky: Kyoto 2006+.}
		\raggedright{\tiny{ * Dataset notation in `Dataset' column. I: ISCX2012, A: AWID, N: NSL-KDD, K: KDDCUP99, U: UNSW-NB15, NB: NIMS botnet, P: Private, AWI: Aegan WiFi Intrusion, Ab: AbdroZoo, D: Drebin, C: CTU-13, Ky: Kyoto 2006+}}
		\resizebox{\textwidth}{!}{\begin{tabular}{lllllllll}
				\hline Ref. & Threat & Type of Threat & IoT Use case & Algo used & \begin{tabular}[c]{@{}l@{}}Feature\\   Extraction\end{tabular} & \begin{tabular}[c]{@{}l@{}}Feature\\   Selection\end{tabular} & Dataset & Accuracy \\ \hline 
				
				Diro et al. \cite{Diro2018} & DoS & Flooding & Fog & LSTM & - & - & I, A & I (99.91), A (98.22) \\
				
				Abeshu et al. \cite{Abeshu2018} & DoS & Flooding & Fog & Softmax & SAE & - & N & 99.2 \\
				
				Tan et al. \cite{Tan2014} & DoS & Flooding & NIDS & TAB & MCA & Norm. & K & normalized 99.95 \\
				
				Tan et al. \cite{Tan2015} & DoS & Flooding & CV & EMD & MCA & PCA & K, I & K (99.95), I(90.12) \\
				
				Moustafa et al. \cite{Moustafa2018} & Botnet & Flooding & IoT & Adaboost & CC & - & U, NB & U(99.54) \\
				
				Ahmad et al. \cite{Ahmad2018} & MiTM & Impersonation & Healthcare & LSTM RNN & NG & - & P & - \\
				
				Aminanto et al. \cite{Aminanto2017} & MiTM & Impersonation & WiFi & ANN & D-FES & - & AWI & 99.92 \\
				
				Chatterjee et al. \cite{Chatterjee2018} & MiTM & Impersonation & RF Comm & ANN & - & - & P & 99.9 \\
				
				Azmoodeh et al. \cite{Azmoodeh2018} & Malware & Code Ijnection & IoBT & DCN & OpCodes & IG & P & 98.37 \\
				
				Aonzo et al. \cite{Aonzo2017} & Malware & Malware & Android & - & \begin{tabular}[c]{@{}l@{}}Static Analysis\\   Technique\end{tabular} & Manual & P & 98.9 \\
				
				Wei et al. \cite{Wei2017} & Malware & Malware & Android & NB, C4.5, kNN & \begin{tabular}[c]{@{}l@{}}Dynamic Analysis\\   technique\end{tabular} & NA & P & {-} \\
				
				Feng et al. \cite{Feng2018} & Malware & Malware & Android & ensemble + LR & Manual & Chi-Square & Ab, D & 98.18 \\
				
				Wang et al. \cite{Wang_DriodEnsemble2018} & Malware & Malware & Android & ensemble & String + structural  & ensemble & Multi-sources & 98.4\\
				
				Maimo et al. \cite{Maimo2018} & Anomaly & Anamoly & 5G & LSTM & Weighted Loss & ASD (DBN+SAE) & C & - \\
				
				Niyaz et al. \cite{Niyaz2016} & Anomaly & Anamoly & NIDS & Softmax & \begin{tabular}[c]{@{}l@{}}SAE using\\   Backpropogation\end{tabular} & - & Ky & 2- 88.39, 5- 79.10 \\
				
				Ambusaidi et al. \cite{Ambusaidi2016} & Anomaly & Anomaly & NIDS & LSSVM & MMIFS & FMIS & K, N, Ky & K 99.95,I 90.12 \\ 
				
				Zhou et al. \cite{Zhou2018} & Dataset & Multiple & IoT & DFEL & - & - & N, U & $>$98.5 \\
				
				Prabavathy et al. \cite{Prabavathy2018} & Dataset & Multiple & Fog & OS-ELM & - & - & N & 97.36 \\ \hline 
				
		\end{tabular}}
		
		\label{tab:SecSolML}

	\end{sidewaystable*}
	
	Anomaly detection is a generic technique where any irregular traffic is flagged as a threat. Several studies \cite{Maimo2018, Niyaz2016, Ambusaidi2016} have attempted to provide secure IDS using ML algorithms. In this regard, an unsupervised DL technique called STL was used by Niyaz et al. \cite{Niyaz2016}, and it was based on SAE and SMR. By using NSL-KDD dataset, the comparison was made using 2-class, 5-class, and 23-class classification, and proved 2-class classification to be better than SMR. A multi-class ML-based classification using Mutual Information (MI) was proposed by Ambusaidi et al. \cite{Ambusaidi2016}. For the linearly dependent variable, Mutual Information Feature Selection (MIFS) with Linear Correlation Coefficient (LLC) was used. For the non-linear dependent variable, the authors used FMIS+MI, made changes to the already existing MIFS algorithm \cite{Press:1992:NRC:148286} and showed their novelty. For the Linear model (Flexible Linear Correlation Coefficient based Feature Selection [FLCFS]), the study modified the existing LLC \cite{Press:1992:NRC:148286} and proposed a new model. An MI can cope with linear as well as non-linear dependents. However, its algorithm can cause redundancy to the classification. Ambusaidi et al. \cite{Ambusaidi2016} chose 'estimator', which relied on estimating the entropies of the given data using average densities from each datum to its k-nearest neighbors. Another reason for this study was that the previous studies had not provided any steps as to how they chose $\beta$. The performance was compared using three different datasets of KDDCUP99, NSL-KDD, and Kyoto 2006+, while the metric performance indicators were Accuracy, DR, FPR, and F-measure. Maimo et al. \cite{Maimo2018} focused on 5G application for anomaly detection based on LSTM.  Features extraction was made from network flows using weighted loss function, while feature reduction was made by using DBN and SAE models because of similar structure (where the prediction can be computed using matrix operations followed by the activation function) \cite{Maimo2018}. After implementing their model using CTU-13 botnet dataset, the authors claimed to have obtained a precision of up to 0.95.
	
	Several studies using ML algorithms as a tool have claimed to reduce cyber-attacks effectively. However, Zhou et al. \cite{Zhou2018} based their proposal \textit{Deep Feature Embedding Learning} (DFEL) on DL because traditional ML algorithms took extra time to train data. The comparison of their proposal using the datasets of NSL-KDD and UNSW-NB15 confirmed the improvement in recall level of Gaussian Naive Bayes classifier from 80.74\% to 98.79\%, apart from the running time of SVM significantly reduced from 67.26 seconds to 6.3 seconds. In another similar study \cite{Prabavathy2018}, the authors claimed that the existing ML algorithms were inefficient for IoT applications and therefore a much faster extreme-learning-machine (ELM) could be used instead \cite{Prabavathy2018}. Furthermore, they found that the existing security approaches for IoT were centralized and cloud-based, and they, in turn, inherited latency and high power consumption. The proposed IDS for IoT used fog computing for implementation in a distributed fashion in two steps. In the first step, attack detection at fog nodes used an online sequential extreme learning machine (OS-ELM) to identify the attacks in the incoming traffic from the IoT virtual clusters. In the second step, these detected threats were summarized and analyzed at a cloud server. The results of the new algorithm showed better accuracy, FRP, and TPR after comparison with the existing NB, ANN, and standard ELM. Furthermore, the experimental results using the Azure cloud also confirmed that the fog-computing-based attack detection was faster than the cloud-computing based attack detection. However, the study did not compare the results with any existing ML/DL based algorithm used for fog-computing.
	
	\subsubsection{Privacy efforts}
	Several privacy-preserving ML algorithms have been proposed, as shown in Table \ref{tab:PrivSolML}.
	Similar to security, privacy is also compromised by a MiTM attack. In this regard, several studies have used ML algorithms to counter different types of MiTM attacks. For example, the study by Xiao et al. \cite{Xiao2016} used game theory--a kind of reinforcement learning, which compared the channel states of the data packets to detect spoofing attacks. The authentication process was formulated as a zero-sum authentication game consisting of the spoofers and the receivers. The threshold was determined by using Nash Equilibrium (NE), implemented over universal software radio peripherals (USPRs), and the performance was then verified via field tests in typical indoor environments. 
	
	As an improvement to their work, Xiao et al. \cite{Xiao2018} applied logistic regression to evaluate the channel model information collected from multiple access points to detect spoofing more accurately. A comparison was made using distributed Frank-Wolfe (dFW)-based and incremental aggregated gradient (IAG)-based authentication to reduce overall communication overhead. IAG-based PHY-layer authentication reduced communication overhead and increased detection accuracy. The results showed improved FAR, DR, and computation costs by using a real-time dataset. In addition to authentication issues, Aksu et al. \cite{Aksu2018} raised an argument concerning the wearable device, for which the previous schemes only focused on user authentication. However, the device being used should also be authenticated. Such devices could act as MiTMs, which might have similar user authentication details. However, in the background, it might leak all the information to the attacker. Wearables could only connect to the more powerful base device via Bluetooth with authentication and encryption. Since the device name and encryption keys could be compromised easily, it was therefore much secure to use hardware-based fingerprinting \cite{Aksu2018}. The proposed framework in \cite{Aksu2018} utilized an inter-packet timing-based timing analysis method based on the Bluetooth classic protocol packets. There were four steps in this framework. The first step captured Bluetooth classic packets. The second step extracted the features. In the third step, using probability distributions, the fingerprints were generated. Moreover, as a final step, the stored fingerprints in step three were compared with any new incoming data from wearable devices, to identify any unknown wearable device. By selecting the best algorithm out of twenty from the training results, the study claimed to achieve an accuracy of 98.5\%.
	
	Data plays a crucial role in training an ML model. For example, we can use patients' historical data to make a predictive decision for any new patient. However, patients are reluctant to share their data due to obvious privacy concerns. The studies, as shown in \cite{Zhu2017, Jia2018, Ma2018}, have worked towards solving these issues. In \cite{Zhu2017}, the researchers proposed a new framework called eDiag, which used non-linear kernel SVM to successfully classify medical information, while preserving user data and service provider's model privacy. Previous studies had used HE techniques, which, according to the study, were not appropriate for online medical prediagnosis. Using their framework, Zhu et al. \cite{Zhu2017} claimed to have achieved a classification accuracy of 94\% without compromising privacy. Similarly, the authors in \cite{Jia2018} classified the privacy issues as \textit{learning-privacy problem} and \textit{model-privacy problem} to protect users' sensitive information and model results, respectively. % The data needs protection as it contains sensitive user information related to \textit{learning-privacy}, while the privacy of the model results as well as \textit{testing data} are related to \textit{model-privacy}. 
	
	Jia et al. \cite{Jia2018} argued that the previous work used either gradient-values instead of real-data, or they assumed that the learning model was private, but the learned model was publicly known, or they used complicated encryption procedures. In comparison to all of these studies, Jia et al. \cite{Jia2018} proposed a uniform Oblivious Evaluation of Multivariate Polynomial (OMPE) model, which did not contain complicated encryption procedures. Their results proved that the classification data and learned models were protected from several privacy attacks. The research in \cite{Jia2018} focused on model-privacy issues. However, the learning-privacy problem was not discussed. This issue was solved by Ma et al. \cite{Ma2018}, who argued that encrypting any user-data by the public key was a widely used privacy-preserving technique but at the cost of key management. To preserve the data privacy, Ma et al. \cite{Ma2018} proposed a cloud-based DL model that worked with multiple keys to attaining privacy of the user data called Privacy-preserving DL Multiple-keys (PDLM). In their proposed model, a service provider (SP) sent encrypted user data to the cloud which performs training of the data without knowing the real data. Their evaluation of the PDLM showed that PDLM had successfully preserved privacy with lower efficiency as compared to the conventional non-private schemes. %cy-preserving mechanism. 
	
	To improve ML algorithms privacy, Sun et al. \cite{Sun2018} proposed an improved version of fully HE that reduced the size and noise of the multiplicative cyphertext by using the re-linearization technique. In their scheme, private hyperplane decision-based classification, private Naive Bayes classification, and private decision tree's comparison were also implemented. In a similar paper, the same authors successfully reduced the user-server iterations to half, without compromising privacy.  
	
	{Social media platforms like Twitter and Facebook have enriched people's lives at the cost of privacy issues. Several companies used blacklisting techniques to filter benign traffic. However, a survey showed that 90\% of the people would fall prey to these attacks before they were blacklisted. To prevent these attacks efficiently, ML algorithms were used. However, these algorithms were inefficient in real-time due to their slower learning rate. In a study, Feng et al. \cite{Feng_MSN2018} proposed a multistage detection framework using DL, where an initial detection occurred at a mobile terminal whose results were then forwarded to the cloud server for further calculation. By using CNN as a classification algorithm, the authors claimed to achieve approximately 91\% utilizing the Sino Weibo dataset. Similarly, the lack of privacy protection mechanisms in a VANET environment was raised by Zhang et al. \cite{Zhang2018}. In VANET, Vehicle nodes tend to learn collaboratively, raising privacy concerns, where a malicious node can obtain sensitive data by inferring from the observed data. A single node has limited computational and memory resources. The solution was presented by using collaborative IDS with distributed ML algorithms and resolving the privacy issues by proposing the concepts of dynamic differential privacy to protect the privacy of a training dataset.}

	\begin{table}[]
		\centering \vspace{-0.2cm}
		\caption{{Existing IoT privacy solutions using machine learning algorithms. Here, ToA means type of attack.} }\vspace{-0.2cm}
		\resizebox{\textwidth}{!}{\begin{tabular}{ccccccc}
				\hline 
				\textbf{Ref.} &
				\textbf{Threat} &
				\textbf{ToA} &
				\textbf{Use Case} &
				\textbf{Algo{rithm}} &
				\textbf{Dataset} &
				\textbf{Accuracy}
				
				\\ \hline 
				Xiao et al. \cite{Xiao2016} & MiTM & Spoof detection & WSN & QL, DQ &  Private & - \\
				
				Xiao et al. \cite{Xiao2018} & MiTM & Spoof detection & MiTMO Landmark & Softmax & Private & - \\
				
				Aksu et al. \cite{Aksu2018} & MiTM & Authentication & Wearable devices & best of 20 & Private & (Precision) 98.5\% \\
				
				Ma et al. \cite{Ma2018} & Data Privacy & Data Leakage & Cloud & SGD  & - & {95\%}\\
				
				Zhang et al. \cite{Zhang2018} & Data Privacy & Inference attack & VANET & LR  & NSL-KDD & - \\
				
				Jia et al. \cite{Jia2018} & Data Privacy & Multiple & Distributed Systems & OMPE  & realworld & {-} \\
				
				Zhu et al. \cite{Zhu2017} & Data Privacy & Multiple & Healthcare & SVM  & realworld & 94\%  \\
				
				Sun et al. \cite{Sun2018} & Data Privacy & Multiple & General & HBD, NB, DT & - & - \\
				
				Feng et al. \cite{Feng_MSN2018} & Anomaly & Spam & MSN & CNN & Sino Weibo & 91.34\% \\ \hline
				
		\end{tabular}}
		\vspace{-0.2cm}
		\label{tab:PrivSolML}
	\end{table}
	\vspace{-0.2cm}
	\subsection{Existing solutions using Blockchain Technology}
	
	Blockchain (BC) is a secure mesh network \cite{Baxter2008}, that is fault-tolerant, transparent, verifiable, and audit-able \cite{Christidis2016}. The frequently used keywords to describe BC benefits are \textit{decentralized, P2P, transparent, trust-less, immutable}. These attributes make a BC more reliable than an untrusted central client-server model. The smart contract is a computer protocol on BC which guarantees the execution of a planned event \cite{Chapron2017}. According to Restuccia et al. \cite{Restuccia2018}, the blockchain guarantees data integrity and validity, making it a suitable solution for protection against data tampering in IoT devices. 
	
	\subsubsection{Security efforts}
	
	%Data Integrity
	Several BC-based solutions for supply-chain, identity management, access management, and IoT were proposed \cite{Kshetri2017}. However, the existing solutions either do not respect the time delay, and cannot be applied to the resource-constrained IoT devices \cite{Machado2018}. {In contrast to that some studies, like \cite{Tapas2018} were only focused on the improvement of time response of an IoT device, rather than their security and privacy.} Machado et al. \cite{Machado2018} offered data integrity for Cyber-Physical Systems (CPS) by splitting their BC architecture into three levels: IoT, Fog, and Cloud. At the first level, the IoT devices in the same domain created trust in each other using Trustful Space-Time Protocol (TSTP), which is based on Proof-of-Trust (PoT). At the Fog level, Proof-of-Luck (PoL) was used to create fault-tolerant IoT data which produces a cryptographic digest for a data audit. The data generated from the first level was hashed using SHA-256 and saved temporarily. After the acknowledgment and consensus were reached, the data was permanently stored at the third level of cloud, which is a public ledger. Other than data integrity, the study also offered key management using time synchronization and the location of the node. HECOPS was used to estimate the node's location via multi-lateration, and TSTP provided clock synchronization. The paper proposed to use multiple consensuses, such as PoT and PoL, but it did not cater to any user privacy issue. Another paper \cite{Liang2017} provided data integrity with the idea of securing data collected from the drone using public BC. DroneChain presented had four modules; \textit{drones}, \textit{control system}, \textit{cloud server}, and \textit{a BC network}. Drones were controlled by the control system, and the data was encrypted and stored using the cloud server on a decentralized BC. The resultant system was trusted and accountable, offered instant data integrity, and had a resilient backend. However, the study used PoW, which was not the best choice for a real-time IoT application like drones. In addition, the work did not offer data provenance and user/data security.
	
	%DoS
	DoS attacks are one of the frequently executing attacks due to their comparatively straightforward implementation and the ever-growing number of insecure digital devices. Due to cheap IoT technologies, hackers can easily control multiple IoT devices to launch an attack. According to \cite{Tselios2017}, the SDN top layer is prone to brute force attacks. Since SDN is controlled by software, it can be {targeted} by injecting malicious applications, and also gives rise to the DoS/DDoS attacks. The earlier methods to prevent DDoS are not compatible with a light-weight multi-standard IoT environment. Other than that, SDN can suffer flooding attacks, saturation attacks, and MiTM attacks due to lack of authentication in the plain-text TCP channel. {Tselios et al. \cite{Tselios2017} argued that BC offered a better solution to protect IoT devices from security attacks and enforced trust between multi-vendor devices, as it was decentralized, fault-tolerant, and tamper-proof}. These valuable BC properties make it { 
		resistant} 
	to data tampering and flooding attacks. However, all of the solutions mentioned above were theoretical ideas as no practical implementation was done. In another paper, Sharma et al. \cite{Sharma2017} improved the security vulnerability in SDN by proposing a distributed SDN architecture for IoT using BC called DistBlockNet. The BC was used to verify, validate, and download the latest flow rule table for the IoT forwarding devices. The proposed DistBlockNet model was compared with the existing solutions, and the results were better in terms of real-time security threat detection and overhead usage. 
	
	%MiTM
	In another study, the researchers highlighted a MiTM security gap in a smart-grid, where any malicious actor could modify user data sent over the {I}nternet \cite{Gao2018}. Secondly, the customers could not audit their costly utility bills, because the current smart-grid was unpredictable, and it did not provide any early warnings to the customer indicating higher energy usage. To avoid the above issues, this study proposed to use cryptographic data transmission using public and private keys for the user ID as well as the smart contract, which was placed on a BC. This technique ensured an immutable, secure, and transparent smart-grid system. However, PoW could be extremely expensive and resource exhausting. 
	
	The study in \cite{Hasan2018} argued that the existing logistics systems were neither transparent nor credible to trace. The existing systems were centralized, relied on multiple TTPs, and focused on a single transporter. Hasan et al. \cite{Hasan2018} proposed a proof of delivery system using BC technique. In their transporter system, the nodes were \textit{seller}, \textit{buyer}, \textit{courier services}, \textit{arbitrator} and \textit{Smart Contract Attestation Authority} (SCAA). The initial agreement was a smart contract that was placed on Inter-Planary File System (IPFS) and was executed once all the parties agreed. The item was transported between several transporters as per the smart contract (maximum three in this paper), which was created every time for the next transporter. Finally, once the buyer has verified and collected the item, the payment is released to the seller. In the case of any rejection (i.e., transaction failure), the \textit{arbitrator} takes over, settles the dispute and redistributes the amount based on the negotiated agreement. This proposed physical-asset-delivery system has inherent BC security against MiTM and DoS attacks. However, the authors have not paid any particular attention to user ID management and data privacy.    
	The study by Gupta et al. \cite{Gupta2018} was a simulation done in OMENT++ on one application scenario where the authors claimed to have tackled Sybil attacks as well as the replay attacks in an IoT network. First of all, they introduced a new layered architecture, which had two more layers in the underlying IoT architecture. They explained their algorithm, idea, and work by comparison in terms of metrics of \textit{Transactions added to the BC per second (Ftx)}, \textit{Blocks added to the BC per second (Fblk)}, and \textit{Memory space utilized (Mmempool)}. 
	
	%Anomaly
	IDS is one of the widely used monitoring devices to detect anomaly traffic behavior. In a study by Golomb et al. \cite{Golomb2018}, the authors argued that the current anomaly IDS were not efficient since the training phase considered only benign traffic. An adversary could exploit this vulnerability by injecting malicious data, which might be regarded as benign. Secondly, the trained model might not be as efficient, since it might be missing some IoT device traffic, which was only event-driven by, for example, a fire alarm. Both of the issues were solved by using a Collaborative IoT Anomaly (CIoTA) Detection using BC technique, where all IoT devices of the same type were trained simultaneously. Since a large number of IoT devices were being trained based on their local data traffic, the chances of an adversarial attack were minimum. Each device would generate a locally trained model which would be collaboratively merged into a globally trained model by using BC technique. The study successfully implemented CIoTA and proved its benefits for eliminating the adversarial attacks. However, the separate block generated for each IoT model would increase the amount of data.
	
	%Multiple
	Along with the research on frequently researched security threats such as Data integrity, MiTM, and DoS, several studies have focused on providing solutions to multiple attacks. Sharma et al. in \cite{Sharma2018a} presented an affordable, secure, and always accessible BC technique for  distributed cloud architecture. The combination of SDN and BC implemented the security of the fog nodes. The study brought the resource extensive tasks closer to the edge of an IoT network, which not only ensured better security but also improved end-to-end transmission delay. The authors further claimed that the model was adaptive based on the encountered threats and attacks, and reduced administrative workload. The main focus of this paper was to provide an architecture based on BC-cloud in fog computing, which was scalable, secure, resilient, and fast. The comparison was made in terms of throughput, response time, and false alarm rate. However, there was no consideration to the data privacy, user ID management, or the key management. Similarly, Sharma et al. in \cite{Sharma2018} claimed that the existing Distributed Mobile Management (DMM) lacked robustness against the security threats due to its centralized architecture. Their proposed scheme based on the BC showed improved latency, delay, and energy consumption, without affecting the existing network layout. However, the study used PoW consensus, which is energy-hungry and offered no user privacy. 
	
	All of the above solutions are mentioned in Table \ref{SecSolBC}, where most of the researchers have focused on using PoW as a consensus algorithm, which is not suitable for a real-time IoT application. Moreover, most of them have not considered user anonymity and data integrity.
	
	\begin{table}[h]
		\centering
		\tabcolsep=0.08cm
		\vspace{-0.2cm}
		\caption{{Taxonomy of existing IoT security solutions using blockchain techniques. Here, U, D, and K mean {\it User security}, {\it Data security}, and {\it Key management}, respectively.}} 
		\vspace{-0.2cm}
		
		\resizebox{\textwidth}{!}{\begin{tabular}{llllllll}
				\hline
				\textbf{Ref.} & 
				{\textbf{Threat}} & 
				{\textbf{Use Case}} & 
				{\textbf{BC used}} & 
				{\textbf{BC type}} & 
				{\textbf{Consensus}} & 
				{\textbf{Security}} &  {\textbf{Weakness}} \\ \hline
				{Machado et al.} \cite{Machado2018}   & Data Integrity & {Cyber Physical System} & Ethereum & Public & PoT + PoL & D/K & {Did not address \textit{U}}   \\ \hline
				{Liang et al.} \cite{Liang2017}  & Data Integrity  & Drone & - & Public & PoW & D/K & {(i) PoW is inefficient for real-time applications}  \\ 
				&   &  &  &  & & & {(ii) Public BC is insecure}  \\ \hline
				{Tselios et al.} \cite{Tselios2017}  & DoS  & SDN & NG & Public & - & None & {\textit{U/D/K} not addressed} \\ \hline
				{Sharma et al.} \cite{Sharma2017}  & DoS  & SDN & Bitcoin & Public & PoW & None & Lack of data integrity \& {\textit{U}}  \\ \hline
				{Gao et al.} \cite{Gao2018}  & MiTM  & SmartGrid & - & Private & PoW & U/D/K & {Encryption techniques are complex and slower} \\ \hline
				{Hasan et al.} \cite{Hasan2018}  & MiTM  & logistics & Ethereum & Private & PoW & K & {Did not address \textit{U} \& \textit{D}}. Overall less secure \\ \hline
				{Gupta et al.} \cite{Gupta2018}  & MiTM  & IoT & Bitcoin & Public & Private & K & Only simulation is done for basic security \\ \hline
				{Golomg et al.} \cite{Golomb2018}  & Anomaly  & Network & Private & Public & Private & D/K & Block per IoT model will increase the data.
				\\ \hline            
				{Sharma et al.} \cite{Sharma2018a}  & Multiple  & Fog-SDN & Ethereum & Public & Proof-of-Service & None & No {\textit{U} or \textit{D}} is offered  \\ \hline
				{Sharma et al.} \cite{Sharma2018}  & Multiple & 5G & Multiple & Both & Multiple & None & PoW is costly, {plus \textit{U/D/K} not addressed}  \\ \hline  			
		\end{tabular}}
		\vspace{-0.6cm}
		\label{SecSolBC}
		
	\end{table}
	\subsubsection{Privacy efforts}

	Privacy is a complicated issue in a BC that can be accomplished, but at the cost of throughput and speed \cite{Christidis2016}. A hacker can identify the patterns of a permissionless BC since all of the transactions happen in public and make an informed decision about the source. BC-based privacy-preserving was proposed by several researchers to solve this issue  \cite{Wang2018, Li2018a, Lu2018, Zhou2018BC, Rahulamathavan2017, Fan2018_Privacy, Aitzhan2018, Guo2018, Kang2017}.
	
	Wang et al. proposed a BC-based model, tackling the MiTM attack issues in a crowdsensing application \cite{Wang2018}. The user privacy was implemented by using node cooperation method, in which the server released the sensing task as well as its price, which was pre-paid on the BC. The users would perform the sensing task and upload the sensing data, and finally, the user was paid as per their achievements. To achieve user-data privacy, the authors proposed k-anonymity, in which the sensing task was not given to an individual, but a group and the sensed data gathered was also in the form of a group, which preserved privacy of a single-user. The announcement VANET is something in which the users (nodes) shared some information that might benefit other users in the network. According to the researchers of CreditCoin \cite{Li2018a}, the current VANET system had a lack of privacy as well as motivation for the users to share any data. CreditCoin was proposed that offered decentralization, trust, and motivation by paying the user their incentives. The shared information was immutable, so the source did not fake any news either, benefiting the whole VANET community from it. For example, the information might be ``a traffic accident on ABC road going towards XYZ''. Another VANET application was proposed by Lu et al. in \cite{Lu2018}, where the authors added privacy to the users in the existing bitcoin platform using the lexicographic Merkle tree. Furthermore, the forgery was controlled by adding a reputation weight to every vehicle in the network.
	However, the study used PoW as their consensus protocol, which is very costly and can create traffic bottlenecks in a resource constraint VANET application.
	
	First, of its nature, Zhou et al. \cite{Zhou2018BC} claimed to design the BC-based IoT system where the servers helped users to process encrypted data without learning from the data. HE was used to secure the data in a private BC using PBFT consensus. The authors in \cite{Rahulamathavan2017} argued that although the BCs were {immutable and tamper-proof}, once a block was executed, they did not cater confidentiality and privacy of the data as anyone could see the plain-text. When such a BC was integrated with IoT, it was more vulnerable due to a massive influx of data. Rahulamathavan et al. focused on these issues by proposing a privacy-preserving BC architecture for IoT applications based on the Attribute-based Encryption (ABE) \cite{Rahulamathavan2017}. 
	
	The previous studies offered the solution by using symmetric encryption like AES, which meant that the key must be shared with the data to enable the miners of the BC to verify the content and update the BC. However, such a technique could not guarantee privacy. ABE used single encryption to keep data private and safe. In a scenario of a hospital, the main server could encrypt data before transmitting the attributes, such as DOCTOR or NURSE, which could only be read by the concerned node by using the same attributes and decrypting them. The BC architecture could secure data manipulation since multiple nodes verified a single transaction. After the approval, the data was stored and could not be tampered. Lastly, there was no central control, making all of the transactions transparent and fair. However, the cluster head could read the data, which might be exploited by an attack. 
	
	Fan et al. working in the 5G network application argued that the work on access control of an encrypted data still needed to be explored \cite{Fan2018_Privacy}. Despite several advantages of ABE, if a user wanted to change his policy, the attribute revocation and re-encryption took much time. Additionally, the owners did not control their public data, and the trust was delegated to the third parties. Centralized systems were fault-prone, and could cause traffic choking. Fan et al. used BC to solve these issues, by using encrypted cloud storage for the provision of privacy-preserving and data-sharing systems, which was tamper-resistant, fully controlled by the user, and always accessible to anyone on request \cite{Fan2018_Privacy}. However, their proposal had several drawbacks; for example, the miners could share the information without user consent. Moreover, the BC proposed is public, which means anyone could access it.
	
	Aitzhan et al. \cite{Aitzhan2018} addressed the issues of transaction security and privacy by using multi-signatures. Since the traditional systems were insecure, unreliable, and publicly accessible, the messages were sent in an encrypted form that offered privacy and security in communication. User anonymity was ensured by using the public key and private key. Similarly, another concept of multi-signatures was mentioned by Guo et al. \cite{Guo2018}. The authors found that the current Electronic Health Record (EHR) system was centralized with no user privacy or control over it. Health records are critical documents as they have a personal medical history. The user should be in control of them, but they should be unforgeable as well. In previous studies, Attribute-Based Signatures (ABS) enabled trust between the two parties; however, it was unreliable and restricted to a single signature. Encashing the ABS advantages, Guo et al. presented an ABS with multiple access (MA-ABS), which guaranteed privacy with access control to the user, and confidence of real information to the verifier \cite{Guo2018}. Moreover, using BC for maintenance of data reinforced immutation, unforgeability, and decentralization. Privacy-preserving was achieved by using MA-ABS and collusion attacks were avoided by using pseudorandom function seed. The study also proposed Key management by using KeyGen. 
	
	In a similar attempt, \cite{Kang2017} offered a new consortium BC called PETCON, that was based on the bitcoin platform using PoW for the PHEV to trade the surplus electricity between them. The existing P2P was a single point of failure, and it was expensive and untrustworthy. Kang et al. \cite{Kang2018} improved upon the privacy of a vehicular data in the existing P2P data sharing networks. Due to the resource constraints in a vehicular system, the data was forwarded to the edge computers for powerful computation. The data shared was vulnerable, due to which, the researchers in this study used consortium BC, where only the selected nodes could perform the audit and verification. They also introduced the use of smart-contracts, which ensured user-authenticity and secure data-sharing, and improved data-credibility. The consortium model reserved the energy as it selected a lesser number of nodes for data maintenance. Vehicle-ID authentication was done by digital signatures using public/private keys, while \textit{Elliptic curve digital signature algorithm} provided key-management. The authors also touched upon data privacy management by storing the raw data using the proof-of-storage.

	\begin{table}[h]
		\centering
		\vspace{-0.2cm}
		\caption{{Overview of existing IoT privacy solutions using blockchain techniques. Here U, D, and K mean {\it User security}, {\it Data security}, and {\it Key management}, respectively.}} 
		\vspace{-0.2cm}
		\resizebox{\textwidth}{!}{\begin{tabular}{llllllll}
				\hline
				{\textbf{Ref.}} & 
				{\textbf{Threat}} & 
				{\textbf{Use Case}} & 
				{\textbf{BC used}} & 
				{\textbf{BC type}} & 
				{\textbf{Consensus}} & 
				{\textbf{Privacy}} & 
				{\textbf{Weakness}}   \\ \hline
				{Wang et al.} \cite{Wang2018}  & MiTM  & Crowdsensing & Bitcoin & Private & PoW & U/D &  {Prone to collusion attacks.}    \\ \hline
				{Li et al.} \cite{Li2018a}  & MiTM  & Vanet & Private & Private & Private & U/D/K & Poor key management     \\ \hline
				{Lu et al.} \cite{Lu2018}  & Data Privacy & VANET & Bitcoin & Private & PoW & U/D/K & PoW is slow \& not ideal for real-time scenario.
				\\ \hline    
				{Zhou et al.} \cite{Zhou2018BC}  & Data Privacy  & IoT & Ehtereum & Private & PBFT & U/D & Block time not suitable for real-time IoT \\ \hline
				{Rahulamathavan et al.} \cite{Rahulamathavan2017}   & Data Privacy  & IoT & Bitcoin & Public & PoW & D/K & Unsuitable for real-time IoT as block time is 10 m.  \\ \hline    
				{Fan et al.} \cite{Fan2018_Privacy} & Data Privacy  & 5G & Private & Public & DPos & U/D/K & Miners can share data \& store data, BC is public. \\ \hline
				{Aitzhan et al.} \cite{Aitzhan2018} & Data Privacy  & Smartgrid & PriWatt & Public &PoC&U& {Did not address \textit{D} and \textit{K}} \\ \hline
				{Guo et al.} \cite{Guo2018}  & Data Privacy  & Healthcare & Private & Public & - & U/D/K & No BC model or consensus technique mentioned. \\ \hline
				{Kang et al.} \cite{Kang2017} & Data Privacy  & PHEV & PETCON & Consortium& PoW & K & {Did not address \textit{U} or \textit{D}} \\ \hline                                
		\end{tabular}}
		\vspace{-0.6cm}
		\label{PrivSolBC}
	\end{table}
	
	\subsection{Existing solutions using Machine Learning and Blockchain}
	
	In this Section, we look at the existing security and privacy solutions for IoT with the integration of ML algorithms and BC techniques.
	
	\subsubsection{Security Solutions}
	
	Agrawal et al. claimed to eliminate spoofing attacks with the combination of ML algorithms and BC techniques \cite{Agrawal2018}. By securing the user-device communication, the user in a valid IoT-zone is continuously monitored, and the communication logs are saved on the BC. The records are immutable and can be verified for any suspicious activities. The existing user authentication techniques include one-time-password (OTP) or security questions, which are limited to single authentication. By using Hyperledger as a BC platform, the authors resolved this issue by considering continuous security using IoT-zone identification, IoT-token generation, and token validation. However, the study considered IoT-hub as a center of communication, which voided the concept of decentralization. There was no user or data privacy in concern, and the dataset was too small for a DL model.
	
	{The open nature of Android poses new security challenges and attacks. Gu et al. \cite{Gu2018} illuminated that Android-based systems were highly targeted by malware, trojans, and ransomware with evolving nature when studied overtime~\cite{zhao2019decade}. The existing schemes, which can be classified as either static-based analysis or dynamic-based analysis, had certain drawbacks such as high computation time costs and types of code obfuscations such as variable encoding and encryption~\cite{IkramBK19}. Gu et al. proposed a new multi-feature detection model (MFM) of Android-based devices, where they utilized a fact-base of malicious codes by using Consortium BC for Malware Detection and Evidence Extraction (CB-MDEE) in mobile devices. Compared with the previous algorithms, CD-MDEE achieved higher accuracy with lower processing time.}
	
	{Using the Exonum BC platform and DNN ML algorithms, the proposed architecture leverage upon BC's properties to send and sell their data \textit{as and when} required giving optimum access control to their health data \cite{Mamoshina2018}. As the data in the storage would be encrypted, the compromise of the storage would not lead to data leakage. The proposed scheme utilizes hash functions and public-key signatures for encrypting user data to guarantee authorization and validity. The paper, however, lacks the in-depth comparison with other schemes, other than being just a theoretical framework.} 
	
	%Existing Security Solutions based on ML and BC integration
	\begin{table*}[h!]
		\centering	
		\vspace{-0.2cm}
		\caption{{Overview of existing IoT security solutions using machine learning algorithms and blockchain techniques. Here, K stands for {\it Key management}.}}
		\vspace{-0.2cm}
		\resizebox{\textwidth}{!}{\begin{tabular}{cccccccccc}
				\hline
				{\textbf{Ref.}} & 
				{\textbf{Attacks}} & 
				{\textbf{Use Case}} & 
				{\textbf{Algo}} & 
				{\textbf{Dataset}} & 
				{\textbf{Metric}} & 
				{\textbf{BC used}} & 
				{\textbf{BC type}} & 
				{\textbf{Consensus}} & 
				{\textbf{Privacy}} \\
				\hline
				
				Agrawal et al. \cite{Agrawal2018}&MiTM&IoT&VMM+ LST&Private&Accuracy&Hyperledger&Private&PBFT&K\\
				\hline
				
				Gu et al. \cite{Gu2018}&Malware&Android&MFM&Drebin&FPR, DR, Acc&Private&Consortium&-&none\\
				\hline
				
				{Mamoshina et al.\cite{Mamoshina2018}} &Access Control & Healthcare & DNN & - & - & Exonum & Private & BFT & U/D/K \\
				\hline
		\end{tabular}}
		\label{table:AllinOneSecurity}
		\vspace{-0.2cm}
	\end{table*}
	
	%Existing Privacy Solutions based on ML and BC integration
	\begin{table*}[h!]
		\centering
		\vspace{-0.2cm}
		\caption{{Summary of existing IoT privacy solutions using machine learning algorithms and blockchain techniques. Here, U, D, and K mean {\it User security}, {\it Data security}, and {\it Key management}, respectively.}}
		\vspace{-0.2cm}
		\resizebox{\textwidth}{!}{\begin{tabular}{cccccccccc}
				\hline
				{\textbf{Ref.}} & 
				{\textbf{Attacks}} &  
				{\textbf{Use Case}} & 
				{\textbf{Algo}} & 
				{\textbf{Dataset}} & 
				{\textbf{Metric}} & 
				{\textbf{BC used}} & 
				{\textbf{BC type}} & 
				{\textbf{Consensus}} & 
				{\textbf{Privacy}} \\
				\hline	
				Mendis et al. \cite{Mendis2018}&Data Leakage& General {IoT} &CNN&Private&Accuracy&Ethereum&Private&PoS&D\\
				\hline
				{Mendis et al. \cite{Mendis2019}}&Data Leakage& SDN & CNN &MNIST&Accuracy&Ethereum&Private&PoS&U/D/K\\
				\hline
				{Weng et al. \cite{Weng2019}}&Data Privacy& General & CNN &MNIST&Accuracy&Corda&Private&BAP\footnotemark[1]&U/D/K\\
				\hline
				Shen et al. \cite{Shen2019} &Data Privacy&Smart Cities&SVM&BCWD+HDD&Accuracy&NG&NG&PoW&U/D/K \\ \hline
				{Goel et al. \cite{Goel2019}} & Data Tampering & Computer Vision & DNN & MNIST/CIFAR-10 & Accuracy & Private & Public & - &U/D/K \\ \hline
				{Fadaeddini et al. \cite{Fadaeddini2020}} & Data Privacy & Self-driving Cars & - & - & - & Stellar & Public & SCP\footnotemark[2] & U/D/K \\ \hline
		\end{tabular}}
		\\
		\centering
		\captionsetup{justification=centering}
		\centering 
		\label{table:AllinOnePrivacy}
		\vspace{-0.6cm}
	\end{table*}
	\footnotetext[1]{Byzantine agreement protocol}
	\footnotetext[2]{Stellar Consensus Protocol}
	
	\subsubsection{Privacy Solutions}
	Many companies rely on big datasets to optimize their target audience and enhance their profits, but such data contain sensitive personal information, such as political preferences, which can be exploited by interested entities. It is, therefore, crucial to preserve the privacy of such users, and if required, compensate them for their contributions. Moreover, certain domains have an abundance of data, which can be beneficial for research and development, but the data cannot be shared with third parties. Furthermore, the same data can be manipulated and raise doubts on its integrity. To improve upon the above architecture, {several studies have been proposed \cite{Meng2018, Mendis2019, Shen2019, Fadaeddini2020, masood2018incognito}.} 
	
	Mendis et al. \cite{Mendis2018} proposed fully autonomous individual contributors working in a decentralized fashion without disturbing the functionality and overall efficiency, {which they later on improved in their work in \cite{Mendis2019}. Their comparison against federated learning using the MNIST dataset for CNN model generated more than 94\% accuracy in each scenario}. The smart contracts incentivizing the computing contributors executed the peer-to-peer transactions. However, {in their study \cite{Mendis2019}, the execution time with encryption increased 100\%. Moreover,} the architecture was based on the ethereum BC having a block-time of 12 seconds, and hence it {might} not feasible for a real-time IoT application, for example, video streaming.
	
	%{DeepChain attempted to solve security issues of federated learning by providing a value-driven incentive mechanism based on BC \cite{Weng2019}. This guarantees data privacy and audit-ability for the model training process. Confidentiality is employed using the Threshold Paillier algorithm that provides an additive homomorphic property. Using CNN algorithms and MNIST dataset, DeepChain proved that the more parties participated in collaborative training, the higher the training accuracy was.}
	{DeepChain proposed BC based value-driven, incentives mechanism to solve security issues~\cite{Weng2019}. DeepChain guarantees data privacy and audit-ability for the model training process. Confidentiality is employed using the Threshold Paillier algorithm that provides an additive homomorphic property. Using CNN algorithms and MNIST dataset, DeepChain proved that the more parties participated in collaborative training, the higher the training accuracy was.}
	
	ML classifiers require datasets to train. These datasets are collected from different entities who are usually reluctant to share their data due to several privacy concerns such as data leakage, data integrity, and ownership. The users do not know how and when their data may be used. To preserve these privacy issues, Shen et al. \cite{Shen2019} proposed a fusion of machine learning with blockchain. A privacy-preserving SVM based classifier was used to train the encrypted data collected from IoT users, while the BC platform provided data sharing among multiple data providers. However, the solution used encryption techniques to preserve privacy, which is not suitable for a resource constraint IoT device. The use of the BC platform is also not explained in detail.
	
	{In yet another study, an attempt to create tamper-proof DNN models is done with the help of BC \cite{Goel2019}. Using the BC properties like \textit{transitive hash}, \textit{cryptographic encryption}, and \textit{decentralized nature}, an architecture named \textit{DeepRing} is proposed. A shared common ledger stored the state of the model. Ouroboros block stored all blocks' hashes, which was used to track the compromised block in case of any tampering attack. Since the querent encrypted the query with its public key, and the output was only encrypted using the public key of the querent, no one else could access the model results. Focusing on the adversarial attacks on network parameters, the authors compared DNN architecture with DeepRing architecture. The DNN architecture without BC using CIFAR-10, MNIST and Tiny ImageNet datasets dropped by their accuracy by 20.71\%, 47\%, and 34\%, respectively. However, the DNN with BC suffered 0\% accuracy loss. }
	
	{Similar work is done in the latest research by Fadaeddini et al. \cite{Fadaeddini2020}, who proposed a framework where the privacy of data-owners was preserved by training the shared model on their data locally. After the learning is completed, the data-owners only shared the learned parameters of the model. The study demonstrated \textit{self-driving cars} application scenario, which used the Stellar BC platform for the decentralized deep learning infrastructure. The contributors are paid for their work as they helped in improving the accuracy of self-driving cars. The learned model is saved on a distributed file system known as IPFS (Inter-Planary File System), which is resistant to DDoS attacks. The framework also controls the authenticity of computing partners to avoid any malicious activities. Although the work is novel and ticks all the privacy issues (i.e., user privacy, data privacy, and key management), however, there is a lack of comparative analysis which can prove that their work is better than the traditional framework.}
	\section{Research Challenges}
	\label{sec:section5:challenges}
	\subsection{Challenge to Machine Learning Algorithms in IoT}
	
	ML algorithms are utilized for analysis after being trained on a large number of datasets to adapt to the desired output dynamically. These models may be used, for example, in navigating a robot or for speech recognition, where human expertise either does not exist or cannot be used. ML algorithms have also been utilized very efficiently to analyze threats against several cybersecurity domains. Although ML algorithms perform well in many areas, they have some limitations in the IoT environment:
	
	\begin{itemize}
		\item \textbf{Scalability and Complexity}:
		In recent studies, several ML algorithms have effectively reduced the cyber attacks. However, ML algorithms are not an ideal pick for IoT applications due to its limitations. Diro et al. claimed that the traditional ML algorithms were limited in scalability, feature extraction, and accuracy \cite{Diro2018}. Whereas, Moustafa et al. \cite{Moustafa2018} argued that ML algorithms could not solve many problems, primarily when it was implemented in a complex resource-constrained IoT environment. Another work done by Abeshu et al. \cite{Abeshu2018} proved that the traditional ML algorithms were less scalable and less accurate in a vast distributed network such as IoT. After comparing classical ML algorithms with DL methods, several studies learned that most DL techniques used pre-training for feature extraction. DL not only saved administrative time but also reduced feature dimensionality by reducing redundancy \cite{Niyaz2016, Kang2016, Sakurada2014, Yan2015, Li2015}. 
		
		\item \textbf{Latency}:
		As a solution to the above issues, some authors, for example, Xiao et al. \cite{Xiao2018} proposed to use ensemble ML algorithms. The ensemble algorithm proved to be performing better than each ML algorithm individually, but it was computationally expensive. As an alternative to classical ML, most of the studies pointed out that DL is a better choice for IoT. In another study, the authors proposed \textit{Deep Feature Embedding Learning} (DFEL) \cite{Zhou2018}. They utilized the DL-based model because the traditional ML algorithms increased training time in Big Data scenarios. Using the datasets of NSL-KDD and UNSW-NB15, they claimed to have improved in the recall of Gaussian Naive Bayes classifier from 80.74\% to 98.79\%. Moreover, their method significantly reduced the running time of SVM from 67.26 seconds to 6.3 seconds. The improvement in recall-rate and running time perfectly suit an IoT application.
		
		\item \textbf{Compatibility}:
		Although the above solutions have performed better, we believe that these DL-based techniques are application-specific. In such cases, a model trained for solving one problem may not be able to perform well for another problem in the similar domain \cite{Hussain2020_MLandIoT}.
		
		\item \textbf{Vulnerability}:
		One of the critical challenges to the ML/DL techniques in IoT is to secure themselves from any security or privacy attacks.  {Adversarial attacks against machine learning models may degrade system performance, as such attacks significantly reduce the output accuracy \cite{Liang2019}. The attack severity is proportional to the amount of information available to an adversary about the system \cite{Chakraborty2018}, which is very difficult to counter. As depicted in Figure \ref{fig:TM4IoT} an adversary can attack ML models at different levels, for example, tampering the input parameters. Goel et al. \cite{Goel2019} highlighted that much work is done to counter input level attacks \cite{Agarwal_extra, akhtar2018threat, Goel2018_Adversarial_Smartbox, Goswami2019_Adversarial, goswami2018unravelling}, however, the research focus on adversarial attacks on network parameters is very less. } Some of these attacks can be proven deadly, for example, in a healthcare application where an ML algorithm is used to analyze the amount of insulin provided by a patient.  
		If an adversary can inject malicious code and alter the ML algorithm's input, the amount of insulin may be increased and cause death to the patient.
		
		Regarding the above issues, we believe that the ML algorithms for IoT {need to be optimised for scalability, speed, compatibility, and security \& privacy.} We think that privacy-preserving ML algorithms, such as differential privacy and light-weight HE, should be explored to overcome the discussed challenges.
		
	\end{itemize}
	\vspace{-0.2cm}
	\subsection{Challenges to Blockchain in IoT}
	
	\begin{itemize} 
		\item \textbf{Latency and speed}:
		Although the BC technology was introduced a decade ago, its real benefits were realized only recently. In recent studies, many efforts have been made to utilize BC in several applications, such as logistics, food, smart grid, VANET, 5G, healthcare, and crowdsensing. However, the existing solutions do not respect the latency issues of BC, and cannot be applied to the resource-constrained IoT devices \cite{Machado2018, Dorri2016}. The most widely used BC consensus is PoW, as depicted in Table \ref{PrivSolBC}. PoW is a slow (limited to seven transactions per second compared to an average of two thousand transactions per second for the visa credit network) and requires a lot of energy \cite{Chapron2017, Christidis2016, Biswas} 
		
		\item \textbf{Computation, processing, and data storage}:
		There is a substantial cost of computation, power, and memory involved in maintaining a BC across a vast network of peers \cite{Song2018, Biswas}. According to the Song et al., in May 2018, the bitcoin ledger size had surpassed 196 GB. These limitations suggest poor scaling and transaction speed for an IoT device. Although an alternative was to offload their computation tasks onto a central server - cloud, or a semi-decentralized server - fog, this, however, adds network latencies \cite{Song2018, Reyna2018}. 
		
		\item \textbf{Compatibility and Standardization}:
		Like any emerging technology, one of the BC challenges is its standardization for which the laws need to be reformed \cite{Niwa2007}. Cybersecurity is a difficult challenge, and it would be naive to think that we all will see a security and privacy standard that can eliminate all risks of cyber-attack against IoT devices anytime soon. Even so, a security standard can ensure that devices meet ``reasonable'' standards for security and privacy. There are a number of fundamental security and privacy capabilities that should be included in any IoT device.

		\item \textbf{Vulnerability}:
		Although the BC is non-repudiable, trustless, decentralized, and {tamper-proof}, a blockchain-based system is only as secure as the system's access point. In a public BC-based system, anyone can access and view the data contents. While {the} private blockchain is one of the solutions to the above problem, it raises other issues {such as} trusted third party, centralized-control, and access-control legislation. {In general, the blockchain-enabled IoT solutions must meet the security and privacy requirements such as (i) the data must be stored securely by satisfying the confidentiality and integrity requirements; (ii) data must be securely transmitted; (iii) data must be shared transparently, securely and in an accountable fashion; (iv) the properties of authenticity and non-reputation must be preserved; (v) the selective disclosure property must be satisfied by the data-sharing platform, and (vi) the explicit consent of data sharing must be taken by the involved parties \cite{ferdous}.}
		
	\end{itemize}
	\vspace{-0.3cm}
	{
		\subsection{Challenges to ML \& BC in IoT}
		We believe that a single technology or a tool, like BC or ML, will not suffice in providing optimum security and privacy for IoT networks. Therefore, it is a dire need of time for the research community to explore the provision of IoT security and privacy with the merger of BC and ML, that has the following challenges:
		\begin{itemize}
			\item \textbf{Storage}: As discussed in Section~\ref{sec:section4:solutions-for-iot}, ML algorithms perform better with larger datasets \cite{Abeshu2018, Diro2018}. However, the increase of data in BC platforms will degrade its performance \cite{Song2018}. It is an open research issue to find a balance, which would be ideal for IoT applications. 
			\item \textbf{Latency challenges:} Depending upon the scenario, an IoT network may generate a considerable amount of data requiring more time for training and computation, which may potentially increase the overall performance (i.e., latency) of traditional ML models~\cite{Dorri2016, Machado2018}. %which will may potentially increase the latency in traditional ML models as it requires more training and computational time. Similarly, the existing BC-based solutions do not respect latency issues in IoT environments \cite{Dorri2016, Machado2018}.
			\item \textbf{Scalability:} ML and BC have scalability challenges, in terms of both the processing and communication costs. Many ML algorithms impose additional processing and communication costs with the increase of data that is imminent for most IoT networks. Similarly, the BC performs poorly as the number of users and networking nodes increases \cite{Salman2018, dinh2017blockbench}. On average, an Ethereum BC performs 12 transactions per second, which is unacceptable in traditional IoT applications, where millions of transactions are happening every second \cite{Salah2019}.
			\item \textbf{Vulnerability:} 
			Although the combination of ML and BC can tremendously increase security and privacy, there are a few challenges as well. The increasing number of threats, including malware and malicious code, increases the challenge of identifying, detecting, and preventing them in real-time IoT networks. The training phase of ML takes longer, and while it is possible to detect malicious traffic, this is only possible with a trained model \cite{Liang2019}. Blockchain, on the other side, can guarantee data immutability and can identify their transformations. However, the issue is with the data that is corrupted before entering the blockchain. Additionally, the malfunctioning of sensors and actuators from the start cannot be detected until that particular device has been tested \cite{Reyna2018}. 
			Besides the above issues, public BC is prone to privacy evasion techniques as the stored data is publicly accessible and available to all readers. Using private BC is one of the solutions to these challenges; however, this would limit access to a large amount of data required for ML to perform efficiently \cite{Salah2019}.
	\end{itemize}}
	
	{The IoT devices can generate a massive amount of data, which should be typically processed in real-time. Since the demand for IoT-based BC is different, there is much research going on to bring a new BC that is compatible with IoT. However, the most important limitations on BC are ledger storage and transaction per second (TPS). Although in the latest BCs, such as Hyperledger Fabric, TPS is down to milliseconds, a lot still needs to be done for a BC to work smoothly in the IoT environment. Similarly, in the context of the secure BC model of IoT, the security needs to be built-in, with validity checks, authentication, and data verification, and all the data needs to be privacy-preserved at all levels. We need a secure, safe, and privacy-preserved IoT framework.} 
	
	%%%%% Conclusion %%%%%
	\vspace{-0.1cm}
	\section{Conclusion and future work}
	\label{sec:section6:conclusion}
	In this paper, we have reviewed the latest threats to IoT and categorized them into security and privacy. Their effects, type of attacks, the layer of impact, and solutions have been briefly mentioned. We have then comprehensively presented the latest existing literature survey on IoT security and privacy using ML algorithms as well as BC technologies and highlighted their gaps. This paper has presented the current solutions to IoT security and privacy by utilizing ML algorithms, BC techniques, and the integration of both. To better understand the security and privacy issues in an ML, we have also attempted to present an ML threat model for IoT based on the previous studies. Finally, We discuss a few research challenges to ML algorithms in IoT, BC techniques in IoT {, and the challenges to the combination of ML and BC in IoT.} 
	
	The generation, storage, analysis, and communication of data are fundamental to the IoT ecosystem. A holistic approach is in demand, where a vulnerability-free system needs to be built, through measures such as adherence to best practices and continual testing. The system should be able to learn and adapt to the latest trends in threats (zero-day attacks) since malicious activities are dynamic. In this regard, ML/DL can be extremely beneficial in analyzing the traffic. At the same time, the BC can serve as a basis to keep a ledger of logs and communication in an IoT environment. Since this data is immutable, it can be used confidently in the court of law as a piece of evidence. 
	
	Among the studies conducted on IoT security and privacy, most of them focused on providing security or privacy. We believe that for a system to be secure, both security and privacy are equally important.  Moreover, data privacy is the most critical factor, which can only be valid when considered end-to-end. The current systems lack the integrity of datasets that are used to train a model. Any adversary can tamper these datasets to obtain their desired results.
	
	Currently, the integration of ML algorithms with BC techniques to achieve IoT security and privacy is a relatively new area, which requires further exploration. However, some of the research questions are: (i) Can we use BC to eliminate DDoS attacks in an IoT network by integrating it with ML algorithms? (ii) Can the resource-constrained IoT device leverage upon BC's inherited encryption to perform in real-time? (iii) Can BC introduce trust in traditional collaborative ML-based IoT Intrusion Detection Systems? Moreover, several organizations, both public and private, rely on the data generated by IoT devices. How can we trust the data, whether \textit{in motion}, or \textit{at rest}? This question becomes more difficult to answer in a centralized cloud-based IoT architecture. We can extract meaningful data from privacy-preserving ML algorithms, whereas BC can offer security and trust.  In the future, we aim to design and develop a privacy-preserving IoT framework, which will offer privacy-preserving data sharing and privacy-preserving data analysis.
	
	{\footnotesize
		\bibliographystyle{ACM-Reference-Format}
		\bibliography{PaperReferences}}
	
\end{document}